\documentclass[12pt]{article}

\usepackage{epsfig}
\usepackage{amsmath}
\usepackage{amssymb}
\usepackage{axodraw}

\begin{document}

\marginparwidth 3cm
\setlength{\hoffset}{-1cm}

\providecommand{\Mink}{\mathbb{R}^{1,9}}
\providecommand{\Ibar}{\bar{I}}
\providecommand{\Jbar}{\bar{J}}
\providecommand{\Kbar}{\bar{K}}
\providecommand{\Lbar}{\bar{L}}
\providecommand{\Mbar}{\bar{M}}
\providecommand{\Nbar}{\bar{N}}
\providecommand{\Lbar}{\bar{L}}
\providecommand{\Abar}{\bar{A}}
\providecommand{\Bbar}{\bar{B}}
\providecommand{\Cbar}{\bar{C}}
\providecommand{\Dbar}{\bar{D}}
\providecommand{\Ebar}{\bar{E}}
\providecommand{\Fbar}{\bar{F}}
\providecommand{\Gbar}{\bar{G}}
\providecommand{\ctil}{\tilde c}
\providecommand{\etil}{\tilde e}
\providecommand{\gtil}{\tilde g}
\providecommand{\psitil}{\tilde \psi}
\providecommand{\Gam}{\Gamma}
\providecommand{\psibar}{\bar{\psi}}
\providecommand{\Psibar}{\bar{\Psi}}
\providecommand{\chibar}{\bar{\chi}}
\providecommand{\zetabar}{\bar{\zeta}}
\providecommand{\lambdabar}{\bar{\lambda}}
\providecommand{\pslash}{\hspace{-1.9mm} \not \!p}
\providecommand{\Pslash}{\hspace{-1.9mm} \not \!P}
\providecommand{\qslash}{\hspace{-1.9mm} \not \!q}
\providecommand{\epsslash}{\hspace{-1.9mm} \not \!\epsilon}
\providecommand{\Prefac}{\frac{1}{(4\pi)^{5/3} \kappa^{1/3}}}
\providecommand{\Ztwo}{\mathbb{Z}_2}
\providecommand{\mydot}{\hspace{-1mm}\cdot\hspace{-1mm}}
\providecommand{\tr}{\text{tr}}

\numberwithin{equation}{section}

\begin{titlepage}

\vspace{1mm}

\begin{flushright}
HUB-EP-99/54
\end{flushright}

\vspace{1mm}

\begin{center}
\baselineskip25pt
{\Large\bf Testing Stability of M-Theory on an $S^1/\Ztwo$ Orbifold}
\end{center}

\vspace{1.1cm}

\begin{center}
\baselineskip12pt
Axel Krause
\vspace{0.3cm}

{\it Humboldt-Universit\"{a}t, Institut f\"{u}r Physik,
     D-10115 Berlin, Germany}

\vspace{1.2cm}
\end{center}

\vspace*{\fill}

\begin{abstract}
We analyse perturbatively, whether a flat background with vanishing
$G$-flux in Ho\v{r}ava-Witten supergravity represents a vacuum state, which
is stable with respect to interactions between the ten-dimensional
boundaries, mediated through the D=11 supergravity bulk fields.
For this, we consider fluctuations in the graviton, gravitino and
3-form around the flat background, which couple to the boundary $E_8$
gauge-supermultiplet.
They give rise to exchange amplitudes or forces between both boundary
fixed-planes. In leading order of the D=11 gravitational coupling
constant $\kappa$, we find an expected trivial vanishing of all three
amplitudes and thereby stability of the flat vacuum in the static limit, in
which
the centre-of-mass energy $\sqrt{s}$ of the gauge-multiplet fields is zero.
For $\sqrt{s}>0$, however, which could be regarded a vacuum state with
excitations on the boundary, the amplitudes neither vanish nor
cancel each other, thus leading to an attractive force between the
fixed-planes in the flat vacuum. A ground state showing stability with regard
to boundary excitations, is therefore expected to exhibit a non-trivial
metric. Ten-dimensional Lorentz-invariance requires a warped geometry.
Finally, we extrapolate the amplitudes to the case of coinciding
boundaries and compare them to the ones resulting from the weakly coupled
$E_8 \times E_8$ heterotic string theory at low energies.
\end{abstract}

\vspace*{\fill}
\end{titlepage}

\section{Introduction and Motivation}
With the discovery of M-theory on an $S^1/\mathbb{Z}_2$-orbifold
\cite{HorWitt1} and its concrete low-energy realization as
Ho\v{r}ava-Witten supergravity \cite{HorWitt2}, i.e.~D=11 supergravity
coupled to two super Yang-Mills theories with $E_8$ gauge group, living on
two separate boundaries of space-time, the vexing
problem of predicting the correct magnitude for the D=4
Newton-constant $G_N$ could be addressed anew.
While the heterotic string theory predicts a value for $G_N$ which is
generically too large by a factor of 400, M-theory on $S^1/\mathbb{Z}_2$
could account for the correct value by adjusting an additional parameter, the
distance $d$ between the two boundaries, roughly at the
inverse of the GUT-scale
$10^{16}$ GeV \cite{Witten}. Similar to the weak-strong coupling
relationship
between the Type IIA-string and
M-Theory, it was found for M-theory on
$S^1/\mathbb{Z}_2$, that $g_s^{2/3} = d/\pi$, where $g_s$ is
the heterotic string coupling constant. Since in the limit
$d\rightarrow 0$ of coinciding
boundaries the string coupling turns out to be weak,
it is believed that we should recover the usual heterotic string theory
with gauge group $E_8\times E_8$. Thus M-theory on $S^1/\mathbb{Z}_2$ has
also been termed heterotic M-theory.
Since even more phenomenological virtues of heterotic M-theory were
discovered, e.g.~it avoids the problem of small gaugino masses
\cite{Nilles}, it is an interesting question to ask for its stability
-- in particular if the boundary fields have non-vanishing energy.

Since the boundaries of the theory maintain an $E_8$ super
Yang-Mills gauge theory, respectively, a non-vanishing energy-momentum
tensor gets induced on each of them. Because gravity couples to any
energy-momentum tensor, an interaction between the boundaries mediated by
gravitons in the bulk is inevitable. This interaction should be
attractive, as can be expected from classical gravity. But furthermore
the D=11 supergravity
bulk theory allows for gravitino and 3-form exchanges, which do
couple to the boundary fields as well, due to the underlying
supersymmetry. Therefore we have to
analyze for heterotic M-theory on the proposed \cite{HorWitt1,HorWitt2}
$\Mink\times S^1/\mathbb{Z}_2$ space-time,
whether all these contributions cancel each other, leading to a stable
configuration or not. Since it is not known how to quantize D=11 supergravity
consistently, we will restrict ourselves to a purely classical
analysis of the stability problem. Remembering the well-known
derivation of the complete Coulomb or Newton potential from tree-level
photon or graviton exchange diagrams, this should not amount to a severe
restriction. Noting that the construction of
Ho\v{r}ava-Witten supergravity has been achieved as an expansion in powers
of small $\kappa^{2/3}$, where $\kappa$ is the D=11 gravitational coupling
constant, we are furthermore allowed to examine the interactions
between the two boundaries
in linearized gravity to leading order in $\kappa$ and discard higher
order contributions as subleading corrections.

One may be inclined to argue that the situation should be similar
to the analogous case of an interaction between two D-branes of Type II
string theory (see \cite{Bachas} for a review). There the repulsion
of the RR-field compensates exactly the attraction originating from
graviton and dilaton exchange. However, in order to reach that conclusion
we have to avail ourselves of the duality between the closed-string
tree-level cylinder amplitude and the open-string 1-loop annulus
diagram. Only through the latter is it possible to see the cancellation
by appealing to Jacobi's {\it aequatio identica satis abstrusa}.
This is in accord with the common lore that supersymmetry leads to
cancellations between fermionic and bosonic loop-contributions
(most prominently applied to the solution of the weak hierarchy problem).
In contrast to the Type II case, heterotic M-theory has been
formulated only as a classical field theory, so far. Therefore, we have
to deal with genuine tree diagrams (without any duality to some possibly
vanishing loop counterpart), for which, even in a supersymmetric
theory, there is {\it a priori} no reason that they add up to zero.
If one could consistently (note that supergravity is non-renormalizable) work
out the Casimir-energy at the one-loop level, then first this could be
expected to vanish on account of the presence of supersymmetry in the bulk.
Second this would constitute only a small quantum correction of order $\hbar$
compared to the leading order tree-level result obtained below. For these
reasons we will not explore the quantum Casimir-energy further in this paper.

It is interesting to consult the supersymmetry variations for the bulk
fields. In heterotic M-theory, the incorporation of $E_8$ super Yang-Mills
theories on the two orbifold fixed-planes, simultaneously requires the 
augmentation of the susy-variations of the bulk-fields \cite{HorWitt2}.
The additional contributions have support on the fixed-planes only and are
solely built out of the boundary-fields.
For the particular flat Minkowski background with vanishing G-flux, which we
will examine
later on, the bulk contributions completely vanish, since flat space does not
break any supersymmetry at all. The only non-vanishing contributions for
constant Majorana-spinor $\eta$ derive from the boundary fields
       \begin{alignat}{3}
              \delta C_{11BC}
            &=-\frac{1}{24\sqrt{2}\pi}\left( \frac{\kappa}{4\pi} \right)^{2/3}
               \delta\left(x_i^{11}-d_i\right)
               \bar{\eta}A^a_{[B}\Gam_{C]}\chi^a_i 
                                            \label{Var1}            \\
              \delta \Psi_A
            &=-\frac{1}{576 \pi}
              \left( \frac{\kappa}{4\pi} \right)^{2/3}
              \delta\left(x_i^{11}-d_i\right)
              \left( \chibar^a_i \Gam_{BCD} \chi^a_i \right)
              \left(\Gam_A^{\phantom{A}BCD}-6\delta_A^B\Gam^{CD}\right)\eta
                                            \label{Var2}            \\
              \delta \Psi_{11}
            &=\frac{1}{576 \pi}
              \left( \frac{\kappa}{4\pi} \right)^{2/3}
              \delta\left(x_i^{11}-d_i\right)
              \left( \chibar^a_i \Gam^{ABC} \chi^a_i \right)
              \Gam_{ABC} \eta               \label{Var3}            \; . 
       \end{alignat}
In momentum-space these contributions will vanish in the case of equal momenta
of the boundary fields. Contracting $\delta C_{11BC}$ with the momentum $p_2^C$
of the gauge-field $A^a_B$, we get an expression proportional to
\begin{equation} 
  \bar{\eta} \big( A^a_B(p_2) \pslash_2-p_2\cdot A^a(p_2)\Gam_B \big)
  \chi^a(p_3) \; .
\end{equation}
Choosing the Lorentz-gauge, the second term disappears, whereas the first term
gives zero, when we choose $p_2=p_3=p$ on account of the massless
Dirac-equation $\,\pslash\chi^a(p)=0$. 
For the last two gravitino-variations, we note that 
\begin{equation}
    \chibar^a(p)\Gam^{ABC}\chi^a(p^\prime) =
   -\chibar^a(p^\prime)\Gam^{ABC}\chi^a(p) \; ,
\end{equation}
from which we easily recognize, that the gaugino bilinear
$\chibar^a\Gam^{ABC}\chi^a$ also vanishes in the limit of coinciding momenta
for $\chibar^a$ and $\chi^a$. In kinematical language, coinciding momenta mean
a vanishing center-of-mass energy squared $s=0$.
Hence, in this limit we expect to find no interaction between the boundaries,
for the assumed flat vacuum.

The interaction amplitudes will depend on the parameter $d$, representing
the distance between the two boundaries in the eleventh direction.
In case that we can still trust Ho\v{r}ava-Witten supergravity not
only for large values of $d$ but also for small values, then
according to the conjecture, the $d\rightarrow 0$ limit of the above
amplitudes should correspond to the low-energy limit of heterotic
string amplitudes.
Consequently we will derive the adequate string expressions
describing an exchange of D=10 supergravity multiplet fields in the
low-energy limit
$\alpha' s,\alpha' t,\alpha' u \ll 1$ and compare them with our M-theory
amplitudes evaluated at $d=0$. Naively, one would not expect complete
agreement of the two sets of amplitudes, since a large $d$ compared to the
eleven-dimensional Planck-scale is a necessary condition for the validity of
the effective Ho\v{r}ava-Witten supergravity.

The rest of the paper is organized as follows. In section 2 we will
apply the
background field method to Ho\v{r}ava-Witten supergravity with
$\Mink\times S^1$ chosen as the background\footnote{We are working
       in the ``upstairs'' formulation \cite{HorWitt2}, regarding the theory
       as an orbifold on $\Mink\times S^1$ with a $\mathbb{Z}_2$ symmetry
       imposed on the fields.}.  
After gauge fixing the relevant symmetries, we obtain the bulk
propagators and determine the leading order couplings of the boundary
fields to the bulk supergravity fields. Section 3 continues with
the calculation of the amplitudes for graviton, gravitino and 3-form
exchange between the boundaries and a subsequent analysis of the
stability of
the theory. In Section 4 we derive the $d\rightarrow 0$ limits of
the amplitudes and compare them with their analogues from low-energy
heterotic string theory. Section 5 finally ends with a summary and a
conclusion.

\section{Expansion of Ho\v{r}ava-Witten Supergravity around
         $\Mink\times S^1$ background}
As advocated in \cite{HorWitt1},\cite{HorWitt2} we should choose
$\Mink\times S^1$ as our D=11 space-time manifold, where the eleventh
coordinate $x^{11}$ is curled up to a circle which we parameterize by
$[-d,d]$ with $d\sim -d$ to be identified. Furthermore we have to impose
the constraint, that the
fields be invariant under the reflection $x^{11} \rightarrow - x^{11}$. This
so-called ``upstairs'' formulation, which we shall employ here, has the
advantage that one can work with a smooth manifold, whereas in the
equivalent
alternative ``downstairs'' formulation one would have to deal with a
bounded manifold $\Mink\times S^1/\Ztwo = \Mink\times[0,d]$ and prescribe
suitable boundary conditions. In the latter approach the boundary is
given by the two codimension one fixed planes of the reflection
map, situated at $x^{11}=0$ and $d$.

The construction of Ho\v{r}ava-Witten supergravity proceeds by a power
series in the expansion parameter $\kappa^{2/3}$. To lowest order, one
starts with the action of N=1, D=11 supergravity \cite{CremmJulSche}
      \begin{alignat}{3}  
         S_{bulk} = \int_{\Mink\times S^1} d^{11} x
                          \frac{\sqrt{-g}}{\kappa^2}
                  \bigg[&-\frac{R}{2}
                        -\frac{1}{2} {\bar{\Psi}}_I \Gam^{IJK} D_J \Psi_K
                        -\frac{1}{2\times 4!} G_{IJKL} G^{IJKL}    \notag \\
                       &-\frac{\sqrt{2}}{192} \left( {\bar{\Psi}}_I 
                                           \Gam^{IJKLMN} \Psi_N
                                          + 12 {\bar{\Psi}}^J \Gam^{KL}
                                           \Psi^M \right) G_{JKLM}  
                                                    \label{BulkAction}    \\
                       &-\frac{\sqrt{2}}{3456}\epsilon^{I_1...I_{11}}
                         C_{I_1I_2I_3} G_{I_4...I_7} G_{I_8...I_{11}}\bigg]
                        +{\cal O}\left( \Psi^4 \right)             \notag
      \end{alignat}
for the bulk
multiplet, consisting
of elfbein $e^{\Ibar}_{\phantom{\Ibar}I}$, gravitino $\Psi_I$
and 3-form $C_{IJK}$. We use $I,\hdots,N / A,\hdots,F$ to represent
D=11/D=10 space-time indices and $\Ibar,\hdots,\Nbar /
\Abar,\hdots,\Fbar$
for their tangent space analogues. Moreover we define $\Psibar_\alpha =
C_{\alpha\beta} \Psi^\beta$, where the real, antisymmetric charge
conjugation matrix $C_{\alpha\beta}$ obeys $C^{\alpha\beta} C_{\beta\gamma}
= \delta^\alpha_\gamma$ (see the appendix for our conventions).
The covariant derivative of the gravitino, the spin connection
$\Omega_{J\Lbar\Mbar}$ and the 4-form field strength $G_{IJKL}$ are
defined as
      \begin{alignat*}{3}
         &D_J \Psi_K 
        = \partial_J \Psi_K + \frac{1}{4}\Omega_{J\Lbar\Mbar}
          \Gam^{\Lbar\Mbar}\Psi_K                                        \\
         &\Omega_{J\Lbar\Mbar} 
        = \frac{1}{2}\left( e_{\Lbar}^{\phantom{L}L}
                                           {\tilde \Omega}_{JL\Mbar}
                                          -e_{\Mbar}^{\phantom{L}L}
                                           {\tilde \Omega}_{JL\Lbar}
                                          -e_{\Lbar}^{\phantom{L}L}
                                           e_{\Mbar}^{\phantom{M}M}
                                           e^{\Jbar}_{\phantom{J}J}
                                           {\tilde \Omega}_{LM\Jbar}
                         \right)
               \; , \quad
              {\tilde \Omega}_{JL\Mbar} 
            = \partial_J e_{\Mbar L} - \partial_L e_{\Mbar J}            \\
         &G_{IJKL} = 4!\, \partial_{[I} C_{JKL]} \; .
      \end{alignat*}
Gravitational
anomalies arise on the D=10 fixed planes. The factorizable
$(\tr R^2)^3, \tr R^2 \tr R^4$ terms can be cured by a Green-Schwarz-like
mechanism whereas the irreducible $\tr R^6$ part of the anomaly
necessitates the introduction of 496 D=10
vector-supermultiplets for compensation. Due to the
$\Ztwo$-symmetry they have to be compartmentalized equally to 248
multiplets on each fixed plane, which singles out the $E_8$ gauge group
for the super Yang-Mills theories on each boundary. Altogether the
D=10 boundary action for the $E_8$ vector supermultiplet\footnote{For
                            $E_8$ the Lie-Algebra index $a$ runs from
                            1 to 248.}, comprising the gauge field $A^a$
and the gaugino $\chi^a$, coupled to the
bulk supergravity in a locally supersymmetric fashion, reads
\cite{HorWitt2} ($i=1,2$)
      \begin{alignat}{3}
         &S_{i,bound}(x^{11}=d_i) 
                    =  \int_{\Mink} d^{10} x_i
                       \frac{1}{(4\pi)^{5/3} \kappa^{4/3}} \sqrt{-g} 
                \bigg[-\frac{1}{4} F^a_{iAB} F^{aAB}_i
                      -\frac{1}{2} {\bar{\chi}}^a_i \Gam^A D_A \chi^a_i 
                                                               \notag \\
                     &-\frac{1}{4} {\bar{\Psi}}_A \Gam^{BC} 
                                             \Gam^A F^a_{iBC} \chi^a_i
                      +{\bar{\chi}}^a_i \Gam^{ABC} \chi^a_i
                       \Big[ \frac{\sqrt2}{48} G_{ABC11}
                            +\frac{1}{32} {\bar{\Psi}}_A \Gam_{BC} \Psi_{11}
                      +\frac{1}{32} {\bar{\Psi}}^D \Gam_{DABC} \Psi_{11}
                                       \label{BoundAction}            \\
                     &+\frac{1}{128} \big( 
                                  3 {\bar{\Psi}}_A \Gam_B \Psi_C 
                                  - {\bar{\Psi}}_A \Gam_{BCD} \Psi^D
                                  -\frac{1}{2}{\bar{\Psi}}_D \Gam_{ABC} 
                                                                \Psi^D
                                  -\frac{13}{6}{\bar{\Psi}}^D \Gam_{DABCE} 
                                                                \Psi^E \big)
                              \Big] \bigg]    
                                    \; ,                     \notag
      \end{alignat}
where $d_1 = 0, d_2 = d$ describe the two fixed plane positions.
The non-abelian field strength $F^a_{iAB}$ and the covariant derivative
for the gaugino are defined as usual as
      \begin{alignat*}{3}
         &F^a_{iAB} 
        = \partial_A A^a_{iB} - \partial_B A^a_{iA} + 
          f^a_{\phantom{a}bc} A^b_{iA} A^c_{iB}                          \\
         &D_A \chi^a_i 
        = \partial_A \chi^a_i + f^a_{\phantom{a}bc} A^b_{iA} \chi^c_i
         +\frac{1}{4}\Omega_{A\Bbar\Cbar}
          \Gam^{\Bbar\Cbar}\chi^a_i \; .
      \end{alignat*}
Actually, on the boundary the 4-form field strength receives an additional
contribution resulting in $G_{ABC11} = 4! \partial_{[A} c_{BC11]} 
-\frac{\kappa^{2/3}}{\sqrt{2}(4\pi)^{5/3}} \delta (x^{11}-d_i) \,\omega_{ABC}
( A_A^a )$, where $\omega_{ABC}$ is a Chern-Simons term depending solely
on the $E_8$ gauge field $A^a_A$. But since the second term does not yield
a coupling to the bulk fields, it is irrelevant for our purposes
and will be neglected in the following. Furthermore the gauginos possess
positive chirality $\Gam^{10}\chi^a = \chi^a$.
The gauge coupling constant $\lambda$ has been eliminated from
(\ref{BoundAction}) by means of the relation\footnote{Originally, the
                relation appeared as $\lambda^2 = 2\pi (4\pi\kappa^2)^{2/3}$
                in \cite{HorWitt2}. We employ a further factor of 2, which
                was found in \cite{Conrad}. Since $\lambda$ will enter
                our analysis only through an overall factor, which is the
                same for all amplitudes, this ambiguity will not have an
                influence on our conclusions.}
$\lambda^2 = 4\pi (4\pi\kappa^2)^{2/3}$. It results from demanding that
on the boundary
the non-vanishing gauge variation of the Chern-Simons term $C\wedge G\wedge G$
should be cancelled by the gauge anomaly for the
D=10 Majorana-Weyl gaugino $\chi^a$. Though gauge invariance cannot be
established at the classical level, it should be valid at the
quantum level, at least. The fixed plane gauge action (\ref{BoundAction})
is the second order term in the power series expansion in $\kappa^{2/3}$.
Unfortunately, in the next higher order infinities arise in the form
of $\delta(0)$ terms occuring in the Lagrangean. Formally, these infinities
cancel in verifying supersymmetry. Nevertheless, to arrive at reliable
results, one is forced to truncate the action at this order consistently.
This will become important below.

From the field-theoretic point of view, we have to look for small
fluctuations of the bulk fields in order to mediate interactions between
the boundary fields. This will be achieved by using the background
field method \cite{DeWitt}, according to which we split the bulk
fields $e^{\Mbar}_{\phantom{\Mbar}M}, \Psi_M, C_{MNP}$ into
a fixed classical background $\etil^{\Mbar}_{\phantom{\Mbar}M},
\psitil_M, \ctil_{MNP}$ and the quantum fields
$f^{\Mbar}_{\phantom{\Mbar}M}, \psi_M, c_{MNP}$, which propagate
on this background
      \begin{alignat*}{3} 
         &e^{\Mbar}_{\phantom{M}M} =  \etil^{\Mbar}_{\phantom{M}M}
                                    + \kappa  f^{\Mbar}_{\phantom{M}M}
          \longrightarrow
          g_{MN} = \gtil_{MN}+\kappa h_{MN}                             \\
         &\Psi_M = \psitil_M + \kappa \psi_M                            \\
         &C_{MNP} = \ctil_{MNP} + \kappa c_{MNP} \; .    
      \end{alignat*}
The action is then expanded around the background fields into a
power series of the quantum fields. The further multiplication with the
D=11 gravitational coupling constant $\kappa$ has been chosen
to give the fluctuations ordinary kinetic terms in the action.
Hence the expansion in quantum fields is also one in powers of
$\kappa$. In the following every index will be raised or lowered by
means of the background elfbein or metric.

With the chosen parameterization of the circle of radius $R = d/ \pi$,
our background $\Mink\times S^1$ is described locally
by a flat elfbein\footnote{For flat space the curved space-time indices
                        $M,N,...$ and the tangent space indices
                        $\Mbar,\Nbar,...$ coincide and need not to be
                        distinguished subsequently.}
      \begin{equation*} 
         \etil^{\Mbar}_{\phantom{M}M} = \delta^{\Mbar}_M \; .
      \end{equation*}
In order not to break spontaneously the D=10 Lorentz symmetry, the
background gravitino field $\psitil_M$ as well as the
background 3-form $\ctil_{MNP}$ must vanish
      \begin{equation*} 
         \psitil_M = \ctil_{MNP} = 0 \; .
      \end{equation*}
As we will point out in the following, it is advantageous (though not
necessary) if the background fields fulfill the equations of motion
as is the case for our flat background.

The expansion of the bulk action then proceeds as
        \begin{alignat*}{3}
             \int d^{11}x \bigg( \;&\frac{1}{\kappa^2}
                                \left[ \text{classical Sugra action} \right]
                                 +\frac{1}{\kappa}
                                \left[ \text{linear in quantum fields
                                             $h,\psi,c$} \right]  \\ 
                                &+\left[ \text{quadratic terms for 
                                                        $h,\psi,c$}
                                                         \right]
                                 + {\cal O}(\kappa) \bigg)   \; .
        \end{alignat*}
For our zero-curvature background the leading term vanishes.
Since the coefficients of $h,\psi,c$ in the $1/\kappa$-term are
precisely
the variational derivatives of the action with respect to the
classical fields, they will vanish when the background satisfies
the equations of motion -- as in our case.
In order to extract the propagators from the quadratic part,
we have, according to the usual Faddeev-Popov procedure,
to fix all the gauge symmetries of the quantum fields and introduce
corresponding ghosts. However, since our analysis is intended to
be classical, i.e.~at tree-level, we can neglect the ghost fields.
N=1, D=11 supergravity possesses four different gauge symmetries, which
are fixed as follows\\[-8mm]
\begin{itemize}
      \item D=11 general coordinate invariance $\longrightarrow$ de Donder
               (harmonic) gauge\footnote{As usual we define $\etil =
                                         \sqrt{-\gtil}$.}
               \begin{equation*}
                      \Delta{\cal L}_{GC} = -\frac{\etil}{2}
                      \big( h_{M\phantom{N};N}^{\phantom{M}N}
                             -\frac{1}{2} h^N_{\phantom{N}N;M} \big)^2
               \end{equation*}
      \item $SO(1,10)$ local Lorentz invariance $\longrightarrow$ symmetric
            gauge
               \begin{equation*}
                      \Delta{\cal L}_{LL} 
                   = -\frac{\etil}{2\kappa^{4/(D-2)}}
                      \big( f_{MN}-f_{NM} \big)^2
                   \rightarrow
                     -\frac{\etil}{2\kappa^{4/9}}
                      \big( f_{MN}-f_{NM} \big)^2
               \end{equation*}
      \item local abelian gauge transformations of $C_{MNP}$
            $\longrightarrow$ Lorentz-like gauge
               \begin{equation*}
                      \Delta{\cal L}_{Ab} = -\frac{3}{2}
                      \big( 3!\partial^M c_{MNP}\big)^2
                                          = -\frac{3}{2} (3!)^2
                      \partial_I c^{IJK} \partial^L c_{LJK}
               \end{equation*}  
      \item local $N=1$ supersymmetry $\longrightarrow$ $\Gam^M \psi_M = 0$
            fixing\footnote{$\zeta$ is a free gauge parameter which we will
                            set later on equal to $-\frac{9}{4}$ for
                            simplicity.} \cite{DasFreedman}
               \begin{equation*}
                      \Delta{\cal L}_{SS} = - \frac{\zeta}{2} \psibar_M
                      \Gam^M \Gam^N \partial_N \Gam^L \psi_L
               \end{equation*}
\end{itemize}
Note that $\Delta{\cal L}_{LL}$ is a purely algebraic term and thus does
not contribute to the propagator. From $e^{\Mbar}_{\phantom{M}M}
e^{\Nbar}_{\phantom{N}N} \eta_{\Mbar\Nbar} = g_{MN}$,
$\etil^{\Mbar}_{\phantom{M}M} \etil^{\Nbar}_{\phantom{N}N}
\eta_{\Mbar\Nbar} = \gtil_{MN}$ and $e^{\Mbar}_{\phantom{M}M} = 
\etil^{\Mbar}_{\phantom{M}M} + \kappa  f^{\Mbar}_{\phantom{M}M}$
plus the above gauge fixing of the Lorentz-symmetry, which gauges the
antisymmetric part of $f_{MN}$ away, we conclude that $h_{MN} = 2 f_{MN}
+ {\cal O}(\kappa)$. Therefore we can express the elfbein
fluctuations in terms of the metric fluctuations.
The quadratic terms of the bulk action then lead to the following
propagators in momentum space, valid for flat, unbounded D=11
Minkowski-space (the compactification of the eleventh coordinate on the
circle will manifest itself, later on,
in a replacement of $p^{11}$ by discrete values $p_m^{11}; m\in
\mathbb{Z}$):\\[-7mm]

\noindent
Graviton $h_{MN}$: 
             \begin{equation}
                 \Delta_{M_1 M_2,N_1 N_2}(P) = -2  \left( \eta_{M_1 N_1}
                                                       \eta_{M_2 N_2}
                                                      +\eta_{M_1 N_2}
                                                       \eta_{M_2 N_1}
                                           -\frac{2}{9}\eta_{M_1 M_2}
                                                       \eta_{N_1 N_2} \right)
                                                       \frac{1}{P^2} \; ,
                               \label{PropGraviton} 
             \end{equation}
Gravitino\footnote{$\alpha,\beta,...$ denote $SO(1,10)$ spinor indices.}
$\psi_M$:
             \begin{alignat}{3}
                 {\Delta_{MN}}^{\alpha\beta}(P) 
                &\equiv i \big({\tilde \Delta}_{MN}
                          \big)^\alpha_{\phantom{\alpha}
                                        \gamma}
                 (P) \left(-C^{\gamma\beta}\right) 
                                  \label{TildeProp}                \\
                &= \frac{i}{9}{\left[ 7
                                                    \eta_{MN} \Pslash
                                                   -\Gam_N \Pslash \Gam_M
                                                   +\left( 4 +
                                                          \frac{9}
                                                               {\zeta}
                                                    \right)
                                                    \frac{P_M \Pslash P_N}
                                                         {P^2}
                               \right]}^\alpha_{\phantom{\alpha}
                                                           \gamma}
                               \left(-C^{\gamma\beta}\right)
                               \frac{1}{P^2}          \notag        \\ 
                 &\stackrel{\zeta=-\frac{9}{4}}{\longrightarrow}
                           \frac{i}{9}{\left[ 7 \eta_{MN} \Pslash
                                              -\Gam_N \Pslash \Gam_M
                                       \right]}^\alpha_{\phantom{\alpha}
                                                        \gamma}
                               \left(-C^{\gamma\beta}\right) 
                               \frac{1}{P^2} \; ,
                             \label{PropGravitino}
             \end{alignat}
3-Form $c_{MNP}$: 
             \begin{equation}
                 \Delta_{M_1 M_2 M_3,N_1 N_2 N_3}(P)
                                      = \frac{1}{(3!)^2}
                                        \frac{ \eta_{M_1[N_1}
                                               \eta_{|M_2|N_2}
                                               \eta_{|M_3|N_3]} }
                                             {P^2} \; , 
                             \label{PropForm}
             \end{equation}
where $P=(p^A,p^{11})$ denotes the eleven-dimensional momentum.

The expansion of the boundary action reads schematically
        \begin{alignat*}{3}
             \int d^{10}x_i \bigg( &\;\frac{1}{\kappa^{4/3}}
                              \left[ \text{pure SYM} \right]
                                   +\frac{1}{\kappa^{1/3}}
                              \left[ \text{bulk-boundary interaction
                                           terms, linear
                                           in $h,\psi,c$} \right] \\
                                &\,+{\cal O}(\kappa^{2/3}) \bigg)   \; .
        \end{alignat*}
\phantom{s}\\[-4mm]
Since the leading $1/\kappa^{4/3}$ contribution does not contain any
bulk quantum field, it cannot contribute to the boundary-boundary interaction
and is therefore of no interest to us.
The $1/\kappa^{1/3}$ terms comprise the relevant interaction terms,
whereas higher order $\kappa^{2/3}$ expressions have to be skipped for
two reasons.
First, $\kappa^{2/3}$ terms would introduce
couplings quadratic in $h,\psi,c$. Either these would finally
lead to loop diagrams
of order $\kappa^{4/3}$, which have to be neglected, since
we restrict ourselves to a tree-level analysis. Or in combination with the
couplings linear in $h,\psi,c$, they would give rise to order
$1$
tree diagrams. These are suppressed by a factor of
$\kappa^{2/3}$ against the
leading diagrams of order $1/\kappa^{2/3}$ and therefore
can be neglected, too.
Second, the boundary action (\ref{BoundAction}) has been constructed only
up to order $1/\kappa^{4/3}$. The next higher order in the power
series expansion in $\kappa^{2/3}$ would involve $1/\kappa^{2/3}$
terms. However, it has been argued in \cite{HorWitt2}, that at this order
expressions containing $\delta(0)$ show up. Therefore, in a consistent
truncation of the theory, we have to skip these higher contributions
altogether. In the expansion around the classical background, the bulk
field fluctuations $h,\psi,c$ come equipped with an additional power of
$\kappa$. Summarizing, a consistent truncation implies throwing away all
bulk-boundary interactions of order $\kappa^{1/3}$ or higher. In
particular, the above $\kappa^{2/3}$ contributions have to be omitted.
Thus, the remaining
$1/\kappa^{1/3}$ interaction terms are given by\footnote{By
                                               $x_1, x_2$ we denote the
                                               D=10 flat coordinates of the
                                               two boundaries, respectively.}
        \begin{alignat*}{3}
             S^{(1)}_{i,bound} (x^{11}=d_i) 
                         =  \frac{1}{(4\pi)^{5/3} \kappa^{1/3}}
                            \int d^{10} x_i
                   \bigg[ &-\frac{1}{8}F^{aCD}_iF^a_{iCD} 
                             h^A_{\phantom{A}A}
                           +\frac{1}{2}F^a_{iAC}F^{a\:C}_{iB} h^{AB}
                                                                          \\
                          &+\frac{1}{8}\chibar^a_i\Gam^A\Gam^{BC}\chi^a_i
                            \partial_{[A} h_{B]C}                     
                           -\frac{1}{16}\chibar^a_i\Gam^A\Gam^{BC}\chi^a_i
                            \partial_{[B} h_{C]A}                   \notag \\
                          &-\frac{1}{4}\psibar_A \Gam^{BC} \Gam^A 
                            \chi^a_i  F^a_{iBC}             
                           +\frac{1}{\sqrt{2}}\chibar^a_i\Gam^{ABC}\chi^a_i
                            \partial_{[A} c_{BC11]}                 \notag
                   \bigg]   \; .        
        \end{alignat*}
It can be read off that we obtain a
5-point vertex $AAAAh$,
two 4-point-vertices $AA\chi\psi$, $AAAh$ and four 3-point vertices
$\chi\chi c$, $A\chi\psi$, $\chi\chi h$, $AAh$. The 5-vertex has a
group-theoretic factor which consists of a sum of terms like
$\sum_{e=1}^{248} (f_{eac} f_{ebd} + f_{ead} f_{ebc})$, where $f_{abc}$
are the $E_8$ structure constants. The 4-vertices
are simply proportional to $f_{abc}$. Since finally
in our amplitudes we will sum over all group indices
of the external boundary fields (we sum over all possible exchange
amplitudes between the two boundaries), the 5- and 4-vertices give no
contribution due to the antisymmetry of the structure constants.
If, therefore, we concentrate merely on the 3-vertices, the relevant
couplings are
        \begin{alignat*}{3}
             S^{(1)}_{i,bound} (x^{11}=d_i) 
                  &= \frac{1}{(4\pi)^{5/3} \kappa^{1/3}}
                     \int d^{10} x_i
                     \bigg[ \frac{1}{2} \Big( -\partial^C A^{a,D}_i 
                                               \partial_{[C} A^a_{iD]} 
                                               \eta_{AB}
                                              +\partial_A A^a_{iC} \partial_B 
                                                A^{a,C}_i           \notag \\
                                              &+\partial_C A^a_{iA} \partial^C 
                                                A^a_{iB}
                                              -2\partial_C A^a_{iA} \partial_B
                                                A^{a,C}_i
                                        \Big) h^{AB}
                           +\frac{1}{8} h_{AC} \partial_B
                            \big( \chibar^a_i \Gam^A\Gam^{BC}
                                   \chi^a_i \big)                   \notag \\
                          &-\frac{1}{2}\psibar_A \Gam^{BC} \Gam^A 
                            \chi^a_i  \partial_B A^a_{iC}         
                           +\frac{1}{\sqrt{2}}\chibar^a_i\Gam^{ABC}\chi^a_i
                            \partial_{[A} c_{BC11]} 
                     \bigg]\; .
        \end{alignat*}
Since every
term comprises exactly two boundary fields, it is convenient for
the later comparison with the string amplitudes to
rescale the super Yang-Mills fields $A^a_A,\chi^a$ which bear mass dimensions
$[A]=1, [\chi]=\frac{3}{2}$ to
        \begin{equation*} B^a_A := 
                         \frac{1}{\left(4\pi\right)^{5/6} \kappa^{2/3}}
                         A^a_A
                         \; , \qquad
                         \lambda^a := 
                         \frac{1}{\left(4\pi\right)^{5/6} \kappa^{2/3}}
                         \chi^a \; .
        \end{equation*}
The fields $B^a_A$ and $\lambda^a$ have D-dimensional
mass dimensions $[B^a_A]=(D-2)/2$ and $[\lambda^a]=(D-1)/2$, i.e.~4 and
$9/2$ for D=10. This rescaling gives a ``canonical'' factor of $\kappa$
for the interaction terms, which eventually read
        \begin{equation}
            S^{(1)}_{i,bound}(x^{11}=d_i) 
                         = \kappa
                           \int_{\Mink} d^{10} x_i
                           \Big( {\cal L}_{i,BBh}
                                +{\cal L}_{i,\lambda\lambda h}
                                +{\cal L}_{i,B\lambda\psi}
                                +{\cal L}_{i,\lambda\lambda c} 
                           \Big)   \; ,                            
                                                 \label{Interaction}
        \end{equation}
where
        \begin{alignat}{3}
             {\cal L}_{i,BBh} 
                            &= \partial_A B^a_{iB}(x_i)
                               \partial_C B^a_{iD}(x_i)
                               h_{EF}(x_i,d_i)
                               F^{ABCDEF}  \; ,
                                                        \label{Int1} \\
             {\cal L}_{i,\lambda\lambda h} 
                            &= -\frac{1}{8} h_{AC}(x_i,d_i)
                                      \partial_B
                                      \big[ \lambda^{a\alpha}_i(x_i)
                                            C_{\alpha\beta}
                                            \left( \Gam^C \eta^{AB}
                                                  -\Gam^B \eta^{AC}
                                            \right)^\beta_{\phantom{\beta}
                                                           \gamma}
                                            \lambda^{a\gamma}_i (x_i) 
                                      \big]   \; ,      
                                                        \label{Int2} \\
             {\cal L}_{i,B\lambda\psi} 
                            &= \frac{1}{2}\psi^\alpha_A(x_i,d_i)
                                         C_{\alpha\beta} 
                                         \left( \Gam^{BC} \Gam^A 
                                         \right)^\beta_{\phantom{\beta}
                                                        \gamma}
                                         \lambda^{a\gamma}_i (x_i)
                                         \partial_B 
                                          B^a_{iC} (x_i) \; , 
                                                        \label{Int3} \\
             {\cal L}_{i,\lambda\lambda c} 
                            &= -\frac{1}{\sqrt{2}}
                                        \lambda^{a\alpha}_i (x_i) 
                                        C_{\alpha\beta}
                                        \left( \Gam^{ABC}
                                        \right)^\beta_{\phantom{\beta}
                                                        \gamma}
                                        \lambda^{a\gamma}_i (x_i)
                                        \partial_{[A} 
                                         c_{BC11]}(x_i,d_i)  \; ,
                                                        \label{Int4}
        \end{alignat}
and
        \begin{equation*}
           F^{ABCDEF} = \frac{1}{2} \eta^{A[D} \eta^{C]B} \eta^{EF}
                                  + \eta^{A[C} \eta^{D]F} \eta^{EB}
                                  + \eta^{A[E} \eta^{D]B} \eta^{CF} \; .
        \end{equation*}

\section{Derivation of the boundary-boundary interaction amplitudes}
In order to incorporate the $\mathbb{Z}_2$ fixed point constraints
and to circumvent ambiguities arising from Feynman diagrams involving
Majorana fermions (which allow for twice as many Wick-contractions as
Dirac fermions do), we
choose not to work with Feynman rules in momentum space directly but
to start with a space-time formulation of the S-matrix on
$\Mink\times S^1$. For a tree-level boundary-boundary interaction
the S-matrix reads
        \begin{alignat}{3}
             S        = -&\frac{1}{2} \kappa^2
                          \int_{\Mink} \hspace{-3mm} d^{10} x_1
                          \oint dx_1^{11} 
                          \int_{\Mink} \hspace{-3mm} d^{10} x_2
                          \oint dx_2^{11}                 \notag \\     
                         &\big\langle f \big| T \Big(
                          :{\cal L}_{1,bound}(x_1,x_1^{11})
                            \delta(x_1^{11}):
                          :{\cal L}_{2,bound}(x_2,x_2^{11})
                            \delta(x_2^{11}-d):
                          \Big) \big| i \big\rangle             
                                                \label{Sgeneral} \\
                      = -&\frac{1}{2} \kappa^2
                          \int d^{10} x_1
                          \int d^{10} x_2
                          \big\langle f \big| T \Big(
                          :{\cal L}_{1,bound}(x_1,0):
                          :{\cal L}_{2,bound}(x_2,d):
                          \Big) \big| i \big\rangle   \; ,  \notag  
        \end{alignat}
where ${\cal L}_{i,bound}$ represents one of the four couplings
${\cal L}_{i,BBh}, {\cal L}_{i,\lambda\lambda h}, {\cal L}_{i,B\lambda\psi},
{\cal L}_{i,\lambda\lambda c}$ given in (\ref{Int1})-(\ref{Int4}). 
From the eleven dimensional perspective the fixed point constraints
enter via delta-function sources which generate flat $p^{11}$-spectra
in momentum space.
Momentum is conserved only along the ten flat directions parallel to the
boundaries, whereas there is no such conservation in the eleventh
compactified direction transverse to the boundaries. This fact is also
well-known from studies of radiation off D-branes \cite{Klebanov},
\cite{Myers}. Therefore the kinematic variables $s,t,u$ of the scattering
process are defined through the ten-dimensional momenta $p_1,p_2$
of the incoming states and $p_3,p_4$ of the outgoing states as follows
        \begin{equation*}
             s=-(p_1+p_2)^2 \; , \qquad
             t=-(p_1-p_3)^2 \; , \qquad
             u=-(p_1-p_4)^2 \; .
        \end{equation*}
As usual D=10 energy-momentum conservation implies for massless states
$s+t+u=0$.
The four 3-vertices (\ref{Int1})-(\ref{Int4}) can be combined into
five different tree-diagrams which we will now consider in detail.

\subsection{Graviton exchange}
The first diagram is depicted in fig.\ref{Picture1} and describes the pure
graviton exchange between the boundary gauge fields. Upon substituting
(\ref{Int1}) into (\ref{Sgeneral}) it
yields the following S-matrix contribution
            \setcounter{figure}{0}
            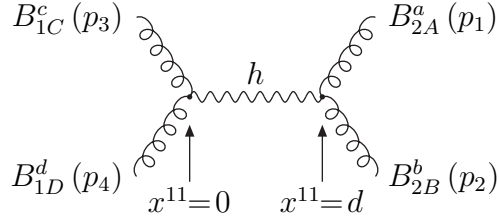
\begin{figure}[t]
              \begin{center}
               \begin{picture}(105,100)(0,0)
                  \Gluon(10,80)(30,50){3}{4}
                  \Gluon(10,20)(30,50){3}{4}
                  \Photon(30,50)(80,50){2}{8}
                  \Gluon(80,50)(100,80){3}{4}
                  \Gluon(80,50)(100,20){3}{4}
                  \Vertex(30,50){1.0}
                  \Vertex(80,50){1.0}

                  \Text(55,55)[b]{$h$}
                  \Text(5,80)[r]{$B^c_{1C} \left(p_3\right)$}
                  \Text(5,20)[r]{$B^d_{1D} \left(p_4\right)$}
                  \Text(105,80)[l]{$B^a_{2A} \left(p_1\right)$}
                  \Text(105,20)[l]{$B^b_{2B} \left(p_2\right)$}
               
                  \LongArrow(30,18)(30,38)
                  \LongArrow(80,18)(80,38)
                  \Text(30,16)[t]{$x^{11}\!\!=\!0$}
                  \Text(80,16)[t]{$x^{11}\!\!=\!d$}
               \end{picture}
               \caption{Graviton exchange}

               \label{Picture1}

              \end{center}
            \end{figure}

       \begin{alignat*}{3}
             S_h
                      = -&\frac{1}{2} \kappa^2
                          \sum_{a,b,c,d = 1}^{248}\;
                          \sum_{\lambda_1,\lambda_2,\lambda_3,\lambda_4=1}^8
                          \int d^{10} x_1
                          \oint d x_1^{11}
                          \int d^{10} x_2
                          \oint d x_2^{11}                              \\
                       &\big\langle 0 \big| 
                              b^d_{1\lambda_4}\left(p_4\right)
                              b^c_{1\lambda_3}\left(p_3\right)
                           T 
                        \Big(
                          :\partial_{A_1} B^{a_1}_{1B_1}\left(x_1\right)
                           \partial_{C_1} B^{a_1}_{1D_1}\left(x_1\right)
                           h_{E_1F_1}\left(x_1,0\right)
                           \delta(x_1^{11}):                            \\
                         &:\partial_{A_2} B^{a_2}_{2B_2}\left(x_2\right)
                           \partial_{C_2} B^{a_2}_{2D_2}\left(x_2\right)
                           h_{E_2F_2}\left(x_2,d\right)\delta(x_2^{11}-d):
                        \Big) 
                              b^{b,\dagger}_{2\lambda_2}\left( p_2 \right)
                              b^{a,\dagger}_{2\lambda_1}\left( p_1 \right)
                           \big| 0 
                        \big\rangle                                     \\
                         &F^{A_1B_1C_1D_1E_1F_1} F^{A_2B_2C_2D_2E_2F_2} 
                         \; .
       \end{alignat*}
Here we sum over all "colours" $a,b,c,d$ and physical polarizations
$\lambda_1,\lambda_2,\lambda_3,\lambda_4$ of the in- and out-states,
since all of them add to the interaction of the two boundaries.
For our conventions concerning annihilation and creation operators
see appendix \ref{Operators}.
When in the next step we Wick-contract creation and annihilation
operators $b^{a,\dagger}_{2\lambda}$ and $b^a_{1\lambda}$
with the boundary field operators $B^a_{iA}$, we have to take into account
that creation and annihilation operators from the left-hand side of the
diagram can only be contracted with left-hand sided $B^a_{1A}(x_1)$
operators and equally creation and annihilation operators from the
right-hand side of the diagram can only be contracted with right-hand sided
$B^a_{2A}(x_2)$ operators. If we would allow for ``mixed'' contractions,
t- and u-channel diagrams would also be present. But these have to be
excluded as they cannot arise when both hyperplanes do not coincide.
After a further trivial integration over the circle coordinates, we are thus
led to
       \begin{alignat*}{3}
             S_h 
                    = -&\frac{1}{2} \kappa^2
                          \sum_{a,b,c,d}\;
                          \sum_{\lambda_1,\lambda_2,\lambda_3,\lambda_4}
                          \int d^{10} x_1
                          \int d^{10} x_2                               \\
                       &\bigg[
                           \thinlines
                           \put(2.4,-5.5){\line(1,0){61.9}}
                           \put(2.4,-5.5){\line(0,1){3}}
                           \put(64.3,-5.5){\line(0,1){3}}
                                b^d_{1\lambda_4}\left(p_4\right)
                               \partial_{A_1} B^{a_1}_{1B_1}
                               \left(x_1\right)
                           \thinlines
                           \put(2.4,-5.5){\line(1,0){61.9}}
                           \put(2.4,-5.5){\line(0,1){3}}
                           \put(64.3,-5.5){\line(0,1){3}}
                                b^c_{1\lambda_3}\left(p_3\right)
                               \partial_{C_1} B^{a_1}_{1D_1}
                               \left(x_1\right)
                          +\big( b^d_{1\lambda_4}\left(p_4\right)
                                  \leftrightarrow
                                  b^c_{1\lambda_3}\left(p_3\right)
                           \big)
                        \bigg]                                          \\
                       &\;\; F^{A_1B_1C_1D_1E_1F_1}
                            i\Delta_{E_1F_1,E_2F_2}
                             \left( x_1-x_2,-d \right) 
                             F^{A_2B_2C_2D_2E_2F_2}                     \\   
                       &\bigg[
                           \thinlines
                           \put(22,-6){\line(1,0){46.5}}
                           \put(22,-6){\line(0,1){3}}
                           \put(68.5,-6){\line(0,1){3}}
                               \partial_{A_2} 
                                B^{a_2}_{2B_2}\left( x_2 \right)
                                b^{b,\dagger}_{2\lambda_2}\left( p_2 \right)
                           \thinlines
                           \put(22,-6){\line(1,0){46.5}}
                           \put(22,-6){\line(0,1){3}}
                           \put(68.5,-6){\line(0,1){3}}
                               \partial_{C_2} 
                                B^{a_2}_{2D_2}\left( x_2 \right)
                                b^{a,\dagger}_{2\lambda_1}\left( p_1 \right)
                          +\big( b^{b,\dagger}_{2\lambda_2}\left(p_2\right)
                                  \leftrightarrow
                                  b^{a,\dagger}_{2\lambda_1}\left(p_1\right)
                           \big)
                        \bigg] \; ,
       \end{alignat*}
where we have expressed the graviton 2-point-function
$\langle 0| T \left( h_{E_1F_1}(x_1,0) h_{E_2F_2}(x_2,d) \right) |0\rangle$
through $i$ times its propagator $\Delta_{M_1M_2,N_1N_2}(x_1-x_2,-d)$.
Using the expressions (\ref{Wick1}) and (\ref{Wick2}) for the
Wick-contractions, gives the $E_8$ group factor
$\sum_{a,b,c,d=1}^{248} \delta^{ab} \delta^{cd} = (248)^2$, and we arrive at
the expression
       \begin{alignat*}{3}
             S_h
                   = -&i\frac{\kappa^2}{2}
                          (248)^2
                          \sum_{\lambda_1,\lambda_2,\lambda_3,\lambda_4}
                          \int d^{10} x_1
                          \int d^{10} x_2 \,
                           e^{i(p_1+p_2)x_2}
                           e^{-i(p_3+p_4)x_1}                           \\
                       &\bigg[ 
                           \epsilon_{B_1}\!\left(p_4,\lambda_4\right)p_{4,A_1}
                           \epsilon_{D_1}\!\left(p_3,\lambda_3\right)p_{3,C_1}
                          +\epsilon_{B_1}\!\left(p_3,\lambda_3\right)p_{3,A_1}
                           \epsilon_{D_1}\!\left(p_4,\lambda_4\right)p_{4,C_1}
                        \bigg]                                          \\
                       &\;\;  F^{A_1B_1C_1D_1E_1F_1}
                              \Delta_{E_1F_1,E_2F_2}
                              \left( x_1-x_2,-d \right)
                              F^{A_2B_2C_2D_2E_2F_2}                    \\   
                       &\bigg[ 
                           \epsilon_{B_2}\!\left(p_2,\lambda_2\right)p_{2,A_2}
                           \epsilon_{D_2}\!\left(p_1,\lambda_1\right)p_{1,C_2}
                          +\epsilon_{B_2}\!\left(p_1,\lambda_1\right)p_{1,A_2}
                           \epsilon_{D_2}\!\left(p_2,\lambda_2\right)p_{2,C_2}
                        \bigg]  \; .
       \end{alignat*}
To utilize the previously derived flat-space propagator
(\ref{PropGraviton}), we have to notice that the momentum in the compactified
eleventh direction $p_m^{11} = m/R ; m 
\in \mathbb{Z}$ is quantized. The radius $R$ of the circle is related via
$R=d/\pi$ to the distance $d$ between the two hyperplanes. Ensuring that we
do not change the dimensions of the propagator as compared to
the flat case, we have to take
       \begin{equation*}
               f\left( x^{11} \right)
            =  \;\frac{1}{2\pi} \int_{-\infty}^\infty d p^{11}
               e^{i p^{11} x^{11}}
               f\left( p^{11} \right)
               \sum_{m \in \mathbb{Z}}
               \delta \left( p^{11} d - m\pi \right)
            =  \;\frac{1}{2\pi d}
               \sum_{m \in \mathbb{Z}}
               e^{i p^{11}_m x^{11}}
               f\left( p^{11}_m \right)
       \end{equation*}
as the Fourier transform in the eleventh direction.
Therefore the Fourier-transformed graviton propagator reads
       \begin{equation*}
               \Delta_{E_1F_1,E_2F_2}\left( x_1-x_2,x^{11} \right)
             = \frac{1}{d \left(2\pi\right)^{11}}
               \int d^{10} p e^{ip\left( x_1-x_2 \right)} 
               \sum_{m\in\mathbb{Z}} e^{ip_m^{11}x^{11}}
               \Delta_{E_1F_1,E_2F_2}\left( p,p_m^{11} \right) \; ,  
       \end{equation*}
where $\Delta_{E_1F_1,E_2F_2}\left( p,p_m^{11} \right)$ is functionally
the same as in the flat, non-compact case. Plugging
       \begin{equation}
               \Delta_{E_1F_1,E_2F_2}\left( x_1-x_2,-d \right)
             = \frac{1}{d \left(2\pi\right)^{11}}
               \int d^{10} p e^{ip\left( x_1-x_2 \right)} 
               \sum_{m\in\mathbb{Z}} \left( -1 \right)^m
               \Delta_{E_1F_1,E_2F_2}\left( p,p_m^{11} \right) \; .
                       \label{SpaceMomProp}  
       \end{equation}
into the expression for $S_h$ and integrating over $x_1, x_2$, results in
       \begin{alignat*}{3}
             S_h 
                    = -&i\frac{\kappa^2}{2 d}
                          \left(248\right)^2 \left(2\pi\right)^{9}
                          \sum_{\lambda_1,\lambda_2,\lambda_3,\lambda_4}
                          \int d^{10} p \;
                          \delta^{10} \left( p_1+p_2-p \right)
                          \delta^{10} \left( -p+p_3+p_4 \right)            \\
                       &\bigg[ 
                           p_{4,A_1}\epsilon_{B_1}\!\left(p_4,\lambda_4\right)
                           p_{3,C_1}\epsilon_{D_1}\!\left(p_3,\lambda_3\right)
                          +p_{3,A_1}\epsilon_{B_1}\!\left(p_3,\lambda_3\right)
                           p_{4,C_1}\epsilon_{D_1}\!\left(p_4,\lambda_4\right)
                        \bigg]                                             \\
                       &\;\;  F^{A_1B_1C_1D_1E_1F_1}
                          \sum_{m \in \mathbb{Z}} \left( -1 \right)^m 
                              \Delta_{E_1F_1,E_2F_2}
                              \left( p , p_m^{11} \right)
                              F^{A_2B_2C_2D_2E_2F_2}                       \\
                       &\bigg[ 
                           p_{2,A_2}\epsilon_{B_2}\!\left(p_2,\lambda_2\right)
                           p_{1,C_2}\epsilon_{D_2}\!\left(p_1,\lambda_1\right)
                          +p_{1,A_2}\epsilon_{B_2}\!\left(p_1,\lambda_1\right)
                           p_{2,C_2}\epsilon_{D_2}\!\left(p_2,\lambda_2\right)
                        \bigg]  \; .
       \end{alignat*}
The integration over $p$ can now be trivially performed, resulting in
an overall D=10 energy-momentum conserving delta-function.
The interaction-amplitude or T-matrix element is defined by equating
$S_h = i \left(2\pi\right)^{10} \delta^{10} \left(
p_1+p_2-p_3-p_4 \right) T_h$.
Going in between to the center-of-mass (CMS) frame with respect to the
ten-dimensional momenta parallel to the boundary, employing
(\ref{PropGraviton}) plus various kinematical relations gathered in
the appendix, we finally arrive at the expression
       \begin{equation*}
             T_h
                    =   \frac{2\kappa^2}{\pi d}
                        \left(248\right)^2
                        \left( 25 s^2 - 32tu \right)  
                        \sum_{m \in \mathbb{Z}} 
                        \frac{\left( -1 \right)^m}
                             {-s + \left(p^{11}_m\right)^2}   \; .
       \end{equation*}
Here $s,t,u$ are the Mandelstam-variables composed out of the
ten-dimensional parallel momenta, as pointed out earlier.
To perform the sum, we use
       \begin{equation*}
             \sum_{m=1}^\infty 
             \frac{\left( -1 \right)^m}
                  {z^2-m^2\pi^2}
           = \frac{1}{2z}\left( \frac{1}{\sin z} - \frac{1}{z} \right)
                \; , \quad  z \in \mathbb{C} 
                \; ,
       \end{equation*}
which one obtains as an application of the Mittag-Leffler theorem
from Complex Analysis
and get
       \begin{equation}
             \sum_{m \in \mathbb{Z}}
             \frac{\left( -1 \right)^m}
                  {-s + \left( p^{11}_m \right)^2}
           = -\frac{d}{\sqrt{s}\sin\left(\sqrt{s}d\right)}   \; .
                               \label{sum}
       \end{equation}
This yields for the matrix-element
       \begin{equation}
             T_h (s, \vartheta)
          = -\frac{2\kappa^2}{\pi}
             \left(248\right)^2
             \frac{ \left( 25 s^2 - 32tu \right) }
                  { \sqrt{s}\sin\left(\sqrt{s}d\right)}
                                      \label{GravitonScattering1}      \; .   
       \end{equation}
It is easy to verify that $25 s^2 - 32tu > 0$.
Concerning our stability analysis, we should further integrate
over the scattering angle $\vartheta$ of the CMS-system from $0$ to
$\pi/2$ (due to the fact that we have identical fields in the out
state, the integration is only over half the usual range). This gives
       \begin{equation}
             {\cal T}_h (s)
          = -21 \left(248\right)^2 \kappa^2
             \frac{ s^{3/2} }
                  {\sin\left(\sqrt{s}d\right)}
                                      \label{GravitonScattering2}      \; .   
       \end{equation}

\subsection{Gravitino exchange}
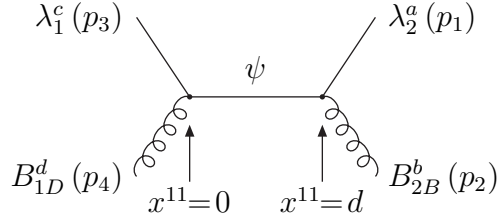
\begin{figure}[t]
              \begin{center}
               \begin{picture}(105,100)(0,0)
                  \Line(10,80)(30,50)
                  \Gluon(10,20)(30,50){3}{4}
                  \Line(30,50)(80,50)
                  \Line(80,50)(100,80)
                  \Gluon(80,50)(100,20){3}{4}
                  \Vertex(30,50){1.0}
                  \Vertex(80,50){1.0}

                  \Text(55,55)[b]{$\psi$}
                  \Text(5,80)[r]{$\lambda^c_1 \left(p_3\right)$}
                  \Text(5,20)[r]{$B^d_{1D} \left(p_4\right)$}
                  \Text(105,80)[l]{$\lambda^a_2 \left(p_1\right)$}
                  \Text(105,20)[l]{$B^b_{2B} \left(p_2\right)$}
               
                  \LongArrow(30,18)(30,38)
                  \LongArrow(80,18)(80,38)
                  \Text(30,16)[t]{$x^{11}\!\!=\!0$}
                  \Text(80,16)[t]{$x^{11}\!\!=\!d$}
               \end{picture}
               \caption{Gravitino exchange, I}

               \label{Picture2}

              \end{center}
            \end{figure}

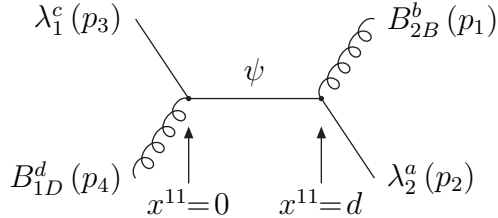
\begin{figure}[b]
              \begin{center}
               \begin{picture}(105,100)(0,0)
                  \Line(10,80)(30,50)
                  \Gluon(10,20)(30,50){3}{4}
                  \Line(30,50)(80,50)
                  \Line(80,50)(100,20)
                  \Gluon(80,50)(100,80){3}{4}
                  \Vertex(30,50){1.0}
                  \Vertex(80,50){1.0}

                  \Text(55,55)[b]{$\psi$}
                  \Text(5,80)[r]{$\lambda^c_1 \left(p_3\right)$}
                  \Text(5,20)[r]{$B^d_{1D} \left(p_4\right)$}
                  \Text(105,20)[l]{$\lambda^a_2 \left(p_2\right)$}
                  \Text(105,80)[l]{$B^b_{2B} \left(p_1\right)$}
               
                  \LongArrow(30,18)(30,38)
                  \LongArrow(80,18)(80,38)
                  \Text(30,16)[t]{$x^{11}\!\!=\!0$}
                  \Text(80,16)[t]{$x^{11}\!\!=\!d$}

               \end{picture}
               \caption{Gravitino exchange, II}

               \label{Picture3}

              \end{center}
            \end{figure}

There exist two diagrams describing amplitudes resulting from gravitino
exchange. The first one is depicted in fig.\ref{Picture2}. Using
(\ref{Int3}) we obtain for its S-matrix\\[-12mm]

       \begin{alignat*}{3}
             S_{\psi,I}
                      = -&\frac{1}{2} \kappa^2
                          \sum_{a,b,c,d = 1}^{248}\;
                          \sum_{\lambda_2,\lambda_4=1}^8
                          \sum_{s_1,s_3=1}^8
                          \int d^{10} x_1
                          \int d^{10} x_2                               \\
                       &\big\langle 0 \big| 
                              b^d_{1\lambda_4}\left(p_4\right)
                              d^c_{1s_3}\left(p_3\right)
                           T 
                        \Big(
                          :\frac{1}{2}\psi^{\alpha_1}_{A_1}(x_1,0)
                           C_{\alpha_1\beta_1}
                           \big( \Gam^{B_1C_1} \Gam^{A_1} 
                           \big)^{\beta_1}_{\phantom{\beta_1}\gamma_1}
                           \lambda_1^{a_1 \gamma_1}(x_1)
                           \partial_{B_1} B^{a_1}_{1 C_1}(x_1):         \\
                         &:\frac{1}{2}\psi^{\alpha_2}_{A_2}(x_2,d)
                           C_{\alpha_2\beta_2}
                           \big( \Gam^{B_2C_2} \Gam^{A_2} 
                           \big)^{\beta_2}_{\phantom{\beta_2}\gamma_2}
                           \lambda_2^{a_2 \gamma_2}(x_2)
                           \partial_{B_2} B^{a_2}_{2 C_2}(x_2):  
                        \Big) 
                              b^{b,\dagger}_{2\lambda_2}(p_2)
                              d^{a,\dagger}_{2 s_1}(p_1)
                           \big| 0 
                        \big\rangle \; .
       \end{alignat*}
Again we sum over "colours" $a,b,c,d$, physical polarizations
$\lambda_2,\lambda_4$ of the gauge fields and spin polarizations $s_1,s_3$
of the gauginos. $d^{a,\dagger}_{2s}$ and $d^a_{1s}$ are the creation
and annihilation operators of the gauginos.
Performing the Wick-contractions with the help of (\ref{Wick1})-(\ref{Wick4})
and using the momentum representation (\ref{SpaceMomProp}) for the
gravitino propagator (\ref{TildeProp}) leads us to
       \begin{alignat*}{3}
             S_{\psi,I} 
                      = &-\left(\frac{1}{2}\right)^2 
                          \frac{\kappa^2}{2d}
                          \frac{(248)^2}{(2\pi)^{11}}
                          \sum_{\lambda_2,\lambda_4=1}^8
                          \sum_{s_1,s_3=1}^8
                          \int d^{10} p \sum_{m \in \mathbb{Z}}(-1)^m
                          \int d^{10} x_1
                          \int d^{10} x_2 
                          e^{-i(-p+p_3+p_4)x_1}                         \\
                         &e^{i(p_1+p_2-p)x_2}
                          p_{4,B_1}\epsilon_{C_1}(p_4,\lambda_4)
                         \bar{u}_{s_3,\gamma_2}(p_3)
                         \big(\Gam^{A_1}\Gam^{B_1C_1}
                         \big)^{\gamma_2}_{\phantom{\alpha_1}\alpha_1}
                         \left({\tilde \Delta}_{A_1A_2}
                         \right)^{\alpha_1}_{\phantom{\alpha_1}\beta_2}
                          (p,p_m^{11})                                  \\
                       &\big( \Gam^{B_2C_2} \Gam^{A_2} 
                           \big)^{\beta_2}_{\phantom{\beta_2}\gamma_2}
                           u^{\gamma_2}_{s_1}(p_1)      
                           p_{2,B_2} \epsilon_{C_2}(p_2,\lambda_2) \; .
       \end{alignat*}
As before the factor of $(248)^2$ represents the $E_8$ group factor.
Performing the $x_1, x_2$ integrations result in two Dirac delta
functions describing the ten-dimensional energy-momentum conservation
at each vertex separately. Upon integration over the momentum $p$,
carried by the gravitino, we arrive at the following T-matrix element
       \begin{alignat*}{3}
             T_{\psi,I} 
                      = &\;i
                          \frac{\kappa^2}{4\pi d}
                          \left(\frac{1}{2}\right)^2 (248)^2
                          \sum_{\lambda_2,\lambda_4=1}^8
                          \sum_{s_1,s_3=1}^8
                          \sum_{m \in \mathbb{Z}}(-1)^m                 \\
                       &p_{4,B_1}\epsilon_{C_1}(p_4,\lambda_4)
                         \bar{u}_{s_3,\gamma_2}(p_3)
                         \big(\Gam^{A_1}\Gam^{B_1C_1}
                         \big)^{\gamma_2}_{\phantom{\alpha_1}\alpha_1}
                         \left({\tilde \Delta}_{A_1A_2}
                         \right)^{\alpha_1}_{\phantom{\alpha_1}\beta_2}
                          (p_1+p_2,p_m^{11})                            \\
                       &\big( \Gam^{B_2C_2} \Gam^{A_2} 
                           \big)^{\beta_2}_{\phantom{\beta_2}\gamma_2}
                           u^{\gamma_2}_{s_1}(p_1)      
                           p_{2,B_2} \epsilon_{C_2}(p_2,\lambda_2) \; .
       \end{alignat*} 
In order to facilitate this expression further, we note that the Weyl
condition $\Gam^{10}u_s(p)=u_s(p), \; \bar{u}_s(p)\Gam^{10} = -\bar{u}_s(p)$
for the gaugino spinor enforces $\bar{u}_s(p)\Gam^{A_1}...
\Gam^{A_{2n}}u_{s'}(p')=0$. Using this observation, the Weyl condition
itself, the Dirac equation $\,\pslash u_s(p) = {\bar u}_s(p)\,
\pslash = 0$ as well as the expression (\ref{PropGravitino}) for
$\tilde{\Delta}_{A_1A_2}$, we receive in the CMS-frame
       \begin{alignat*}{3}
             T_{\psi,I}
                      = &\;i
                          \frac{\kappa^2}{4\pi d}
                          \left(\frac{1}{2}\right)^2 (248)^2
                          \sum_{\lambda_2,\lambda_4=1}^8
                          \sum_{s_1,s_3=1}^8
                          \sum_{m \in \mathbb{Z}}
                          \frac{(-1)^m}{-s+(p_m^{11})^2}                 \\
                       &\bar{u}_{s_3}(p_3)
                          \bigg( E^2\big( 4\cos\vartheta
                                         +28\big)\pslash_4
                                +2E\sin\vartheta\pslash_4
                                 \epsslash(p_2,\lambda_2)\pslash_2
                                -2E^3 \sin\vartheta 
                                 \epsslash(p_4,\lambda_4)               \\
                               &+\frac{2E^2}{3}\epsslash(p_4,\lambda_4)
                                 \pslash_4 \epsslash(p_2,\lambda_2)
                                +E^2\big(\cos\vartheta-\frac{1}{3}\big)
                                 \epsslash(p_4,\lambda_4)
                                 \epsslash(p_2,\lambda_2)\pslash_2
                          \bigg)u_{s_1}(p_1) \; .
       \end{alignat*}
$E=\sqrt{s}$ and $\vartheta$ denote the ten-dimensional CMS-energy and the
scattering angle in the CMS-frame along the hyperplanes
(see \ref{AppMandelstam}).
Employing the explicit expressions for $\bar{u}_{s_3}(p_3),u_{s_1}(p_1)$ and
for the $\Gam$-matrices from the appendix, we get
        \begin{equation*}
             T_{\psi,I} 
                      = -i
                          \frac{\kappa^2}{4\pi d}
                          \left(\frac{1}{2}\right)^2 (248)^2
                          \sum_{m \in \mathbb{Z}}(-1)^m 
                        128 ( s-u )
                          \sqrt{\frac{-u}{s}}
                          \frac{s}{-s+\left(p_m^{11}\right)^2} \; .
        \end{equation*}
In addition to the diagram of fig.\ref{Picture2}, we also have to add
the diagram of fig.\ref{Picture3}
which merely amounts to the exchange of $p_1 \leftrightarrow p_2$ or
$t \leftrightarrow u$ in the preceding diagram. Adding up both
contributions results in 
        \begin{equation*}
             T_\psi 
                      = -i
                          \frac{\kappa^2}{4\pi d}
                          \left(\frac{1}{2}\right)^2 (248)^2
                          \sum_{m \in \mathbb{Z}}(-1)^m 
                           128
                          \bigg( (s-u)
                                 \sqrt{\frac{-u}{s}}
                                +(s-t)
                                 \sqrt{\frac{-t}{s}}
                          \bigg)
                          \frac{s}{-s+\left(p_m^{11}\right)^2} \; .
        \end{equation*}
\phantom{s}\\[-8mm]
If we utilize (\ref{sum}) again, we conclude
        \begin{equation}
             T_\psi (s,\vartheta)
                      = i \frac{8\kappa^2}{\pi} (248)^2
                           \frac{\left( \left( s-u \right)
                                        \sqrt{-u}
                                       +\left( s-t \right)
                                        \sqrt{-t}
                                 \right)}
                                {\sin\left(\sqrt{s}d\right)} \; .
                         \label{GravitinoFull}
        \end{equation}
For the stability analysis we perform a further integration over the
scattering angle $\vartheta$ from $0$ to $\pi$ (as appropriate for
distinguishable fields in the out state), which finally  gives
        \begin{equation*}
             {\cal T}_\psi (s)
                      = i \frac{160\kappa^2}{3\pi} (248)^2             
                        \frac{s^{3/2}}
                             {\sin\left(\sqrt{s}d\right)} \; .
        \end{equation*}

\subsection{3-Form exchange}
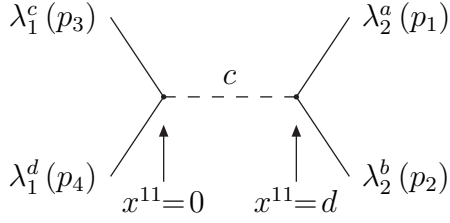
\begin{figure}[t]
              \begin{center}
               \begin{picture}(105,100)(0,0)
                  \Line(10,80)(30,50)
                  \Line(10,20)(30,50)
                  \DashLine(30,50)(80,50){5}
                  \Line(80,50)(100,80)
                  \Line(80,50)(100,20)
                  \Vertex(30,50){1.0}
                  \Vertex(80,50){1.0}

                  \Text(55,55)[b]{$c$}
                  \Text(5,80)[r]{$\lambda^c_1 \left(p_3\right)$}
                  \Text(5,20)[r]{$\lambda^d_1 \left(p_4\right)$}
                  \Text(105,80)[l]{$\lambda^a_2 \left(p_1\right)$}
                  \Text(105,20)[l]{$\lambda^b_2 \left(p_2\right)$}
               
                  \LongArrow(30,18)(30,38)
                  \LongArrow(80,18)(80,38)
                  \Text(30,16)[t]{$x^{11}\!\!=\!0$}
                  \Text(80,16)[t]{$x^{11}\!\!=\!d$}
               \end{picture}
               \caption{3-Form exchange}

               \label{Picture4}

              \end{center}
            \end{figure}

The 3-form exchange diagram of fig.\ref{Picture4} yields the
following expression for the S-matrix element\\[-13mm]

       \begin{alignat*}{3}
             S_c
                      = -&\frac{1}{2} \kappa^2
                          \sum_{a,b,c,d = 1}^{248}\;
                          \sum_{s_1,s_2,s_3,s_4=1}^8
                          \int d^{10} x_1
                          \int d^{10} x_2                               \\
                       &\big\langle 0 \big| 
                              d^d_{1 s_4}(p_4)
                              d^c_{1 s_3}(p_3)
                           T 
                        \Big(
                          :\frac{1}{\sqrt{2}}\lambda_1^{a_1\alpha_1}(x_1)
                           C_{\alpha_1\beta_1}
                           \big( \Gam^{A_1B_1C_1}
                           \big)^{\beta_1}_{\phantom{\beta_1}\gamma_1}
                           \lambda_1^{a_1 \gamma_1}(x_1)
                           \partial_{[A_1} c_{B_1C_1 11]}(x_1,0):       \\
                         &:\frac{1}{\sqrt{2}}\lambda_2^{a_2\alpha_2}(x_2)
                           C_{\alpha_2\beta_2}
                           \big( \Gam^{A_2B_2C_2}
                           \big)^{\beta_2}_{\phantom{\beta_2}\gamma_2}
                           \lambda_2^{a_2 \gamma_2}(x_2)
                           \partial_{[A_2} c_{B_2C_2 11]}(x_2,d):
                        \Big) 
                              d^{b,\dagger}_{2 s_2}(p_2)
                              d^{a,\dagger}_{2 s_1}(p_1)
                           \big| 0 
                        \big\rangle   \; .      
       \end{alignat*}
We make use of (\ref{Wick3}),(\ref{Wick4}) to perform the Wick-contractions
and gain the $E_8$ gauge group factor $(248)^2$ as previously.
We then combine the four resulting terms together by employing
the relation $u^\alpha_s(p)
C_{\alpha\beta}\left( \Gam^{ABC} \right)^\beta_{\phantom{\beta}
\gamma} u^\gamma_{s'}(p') = -u^\alpha_{s'}(p') C_{\alpha\beta}
\left( \Gam^{ABC} \right)^\beta_{\phantom{\beta}\gamma} u^\gamma_s(p)$.
Moreover, expressing the 2-point-function $\langle 0| T
\left( c_{M_1M_2M_3}(x_1,0) c_{N_1N_2N_3}(x_2,d) \right) |0 \rangle$ as
$i$ times the 3-form propagator $\Delta_{M_1M_2,N_1N_2}\left( x_1-x_2,-d
\right)$ gives
       \begin{alignat*}{3}
             S_c 
                      = -&i \kappa^2 (248)^2
                          \sum_{s_1,s_2,s_3,s_4=1}^8
                          \int d^{10} x_1
                          \int d^{10} x_2
                          e^{-i(p_3+p_4)x_1}
                          e^{i(p_1+p_2)x_2}                             \\
                         &u^{\alpha_1}_{s_4}(p_4)
                          C_{\alpha_1\beta_1}
                          \left( \Gam^{A_1B_1C_1} 
                          \right)^{\beta_1}_{\phantom{\beta_1}\gamma_1}
                           u^{\gamma_1}_{s_3}(p_3)
                          \partial_{[A_1} \partial^{[A_2}
                          \Delta_{B_1C_1 11]}^{\phantom{B_1C_1 11]}
                           B_2C_2 11]} (x_1-x_2,-d)                     \\
                         &u^{\alpha_2}_{s_2}(p_2)
                          C_{\alpha_2\beta_2}
                          \left( \Gam^{A_2B_2C_2} 
                          \right)^{\beta_2}_{\phantom{\beta_2}\gamma_2}
                           u^{\gamma_2}_{s_1}(p_1)         \; .
       \end{alignat*} 
Fourier-transforming the propagator with the help of (\ref{PropForm})
and (\ref{SpaceMomProp})
       \begin{alignat*}{3}
                    &\partial_{[A_1} \partial^{[A_2}
                     \Delta_{B_1C_1 11]}^{\phantom{B_1C_1 11]}
                      B_2C_2 11]} (x_1-x_2,-d)                         \\
                   = -&\frac{1}{d(3!)^2(2\pi)^{11}}
                      \int d^{10} p
                      e^{ip(x_1-x_2)}
                      \sum_{m \in \mathbb{Z}} (-1)^m  
                                 \left(\frac{3!}{4!}
                                 \right)^2 
                                 \left( 3p_{[A_1} p^{[A_2} 
                                 \delta^{B_2}_{B_1} \delta^{C_2]}_{C_1]}
                                -(p_m^{11})^2 \delta^{[A_2}_{[A_1} 
                                 \delta^{B_2}_{B_1} \delta^{C_2]}_{C_1]}
                                 \right)                               \\
                      &\phantom{\frac{1}{d(3!)^2(2\pi)^{11}}\int}
                       \times\frac{1}{p^2+(p_m^{11})^2} 
       \end{alignat*} 
brings us to
       \begin{alignat*}{3}
             S_c 
                      =  &\,i \frac{\kappa^2}{d}
                          \frac{(248)^2}{(4!)^2 (2\pi)^{11}}
                          \sum_{s_1,s_2,s_3,s_4=1}^8
                          \int d^{10} p 
                          \sum_{m \in \mathbb{Z}} (-1)^m
                          \int d^{10} x_1
                          \int d^{10} x_2
                          e^{-i(-p+p_3+p_4)x_1}
                          e^{i(p_1+p_2-p)x_2}                             \\
                         &\bigg( 3 u^{\alpha_1}_{s_4}(p_4)
                                 C_{\alpha_1\beta_1}
                                 \left( \Gam^{A_1BC} 
                                 \right)^{\beta_1}_{\phantom{\beta_1}\gamma_1}
                                 u^{\gamma_1}_{s_3}(p_3)
                                 u^{\alpha_2}_{s_2}(p_2)
                                 C_{\alpha_2\beta_2}
                                 \left( \Gam_{A_2BC} 
                                 \right)^{\beta_2}_{\phantom{\beta_2}\gamma_2}
                                 u^{\gamma_2}_{s_1}(p_1)
                                 p_{A_1} p^{A_2}                         \\
                         &- u^{\alpha_1}_{s_4}(p_4)
                                 C_{\alpha_1\beta_1}
                                 \left( \Gam^{ABC} 
                                 \right)^{\beta_1}_{\phantom{\beta_1}\gamma_1}
                                 u^{\gamma_1}_{s_3}(p_3)
                                 u^{\alpha_2}_{s_2}(p_2)
                                 C_{\alpha_2\beta_2}
                                 \left( \Gam_{ABC} 
                                 \right)^{\beta_2}_{\phantom{\beta_2}\gamma_2}
                                 u^{\gamma_2}_{s_1}(p_1)
                                 (p_m^{11})^2
                          \bigg) \frac{1}{p^2+(p_m^{11})^2}   \; .
       \end{alignat*}        
After integration over $x_1, x_2$ and afterwards over $p$,
we recognize the T-matrix element as
       \begin{alignat*}{3}
             T_c = &\frac{\kappa^2}{\pi d}
                          \frac{(248)^2}{(3!)^3} 
                          \sum_{s_1,s_2,s_3,s_4=1}^8
                          \sum_{m \in \mathbb{Z}} (-1)^m                 \\
                         &\bigg( 3 {\bar u}_{s_4}(p_4)
                                 \Gam^{A_1BC} 
                                 u_{s_3}(p_3)
                                 {\bar u}_{s_2}(p_2)
                                 \Gam_{A_2BC} 
                                 u_{s_1}(p_1)
                                 (p_3+p_4)_{A_1} (p_1+p_2)^{A_2}         \\
                         &-      {\bar u}_{s_4}(p_4)
                                 \Gam^{ABC} 
                                 u_{s_3}(p_3)
                                 {\bar u}_{s_2}(p_2)
                                 \Gam_{ABC} 
                                 u_{s_1}(p_1)
                                 (p_m^{11})^2
                          \bigg) \frac{1}{-s+(p_m^{11})^2}   \; .
       \end{alignat*}
In the following, we are dealing separately with the first and the
second term of this amplitude in order to boil them down to some
more enlightening expressions.

Let's start with the first term and the observation that
the Dirac equation $\,{\pslash}\hspace{0.5mm}
 u_s(p) = {\bar u}(p)\hspace{1.4mm}{\pslash} = 0$ gives the relations
\begin{equation*}
   p_A \Gam^{ABC} u_s(p) = 2 p^{[B} \Gam^{C]} u_s(p) \; , \qquad\quad
   {\bar u}_s(p) \Gam^{ABC} p_A = -2 {\bar u}_s(p) p^{[B} \Gam^{C]} \;.
\end{equation*}
If we apply them to the first term, we obtain
        \begin{alignat*}{3}
                         &\sum_{s_1,s_2,s_3,s_4=1}^8
                               3 {\bar u}_{s_4}(p_4)
                                 \Gam^{A_1BC} 
                                 u_{s_3}(p_3)
                                 {\bar u}_{s_2}(p_2)
                                 \Gam_{A_2BC} 
                                 u_{s_1}(p_1)
                                 (p_3+p_4)_{A_1} (p_1+p_2)^{A_2}         \\
                         &= 12
                               \sum_{s_1,s_2,s_3,s_4=1}^8
                                {\bar u}_{s_4}(p_4)
                                 (p_3-p_4)^{[B} \Gam^{C]} 
                                 u_{s_3}(p_3)
                                 {\bar u}_{s_2}(p_2)
                                 (p_1-p_2)_{[B} \Gam_{C]}
                                 u_{s_1}(p_1)                   \; .
       \end{alignat*}  
By noticing that in the CMS-frame $p_1-p_2 = (0,...,0,E)$ and
$p_3-p_4 = (0,...,0,E\sin\vartheta,$ $E\cos\vartheta)$, we eventually reduce
this expression to
       \begin{alignat*}{3}
                              &4! E^2
                               \sum_{s_1,s_2,s_3,s_4=1}^8
                               \big( \cos\vartheta {\bar u}_{s_4}(p_4)
                                     \Gam^9 u_{s_3}(p_3)
                                     {\bar u}_{s_2}(p_2) \Gam_9
                                      u_{s_1}(p_1)                     \\
                              &\phantom{\big)}
                                    +\sin\vartheta {\bar u}_{s_4}(p_4)
                                     \Gam^9 u_{s_3}(p_3)
                                     {\bar u}_{s_2}(p_2) \Gam_8
                                      u_{s_1}(p_1)
                                    -\cos\vartheta {\bar u}_{s_4}(p_4)
                                     \Gam^A u_{s_3}(p_3)
                                     {\bar u}_{s_2}(p_2) \Gam_A
                                      u_{s_1}(p_1)
                               \big)                               \\[2mm]
                           = -&4! \times 64 s^2    \; .
       \end{alignat*}

Concerning the second part of the amplitude, we decompose
       \begin{equation*}
           \Gam^{ABC} = \left( \begin{array}{cc} a & b \\
                                                 c & d
                               \end{array} 
                        \right) 
          \otimes       \left( \begin{array}{cc} A & B \\
                                                 C & D
                               \end{array} 
                        \right)
       \end{equation*}
into $2\times 2$ and $16\times 16$ matrices. With the explicit
expression for the gaugino-spinor (\ref{GauginoSpinor}), we find
       \begin{alignat*}{3}
                     &{\bar u}_{s_4}(p_4)
                      \Gam^{ABC} 
                       u_{s_3}(p_3)
                     = E c \big( \frac{\sin\vartheta}{2} A
                                +\cos^2\frac{\vartheta}{2} B 
                                -\sin^2\frac{\vartheta}{2} C
                                -\frac{\sin\vartheta}{2} D
                           \big)_{s_4s_3}                          \\
                     &{\bar u}_{s_2}(p_2)
                      \Gam^{ABC} 
                       u_{s_1}(p_1)
                     = E c B_{s_2s_1}  \; .    
       \end{alignat*}
In particular, we have for our chosen representation (given in
\ref{AppGamma}) of $SO(1,10)$
$\Gam$-matrices the following description in terms of $8\times 8$
submatrices $\gamma^a$
       \begin{alignat*}{3}
                   &\Gam^{0ab}:\quad c=1, B=0 \qquad\qquad\quad
                    \Gam^{0a9}:\quad c=1, B=\gamma^a \qquad\quad   \\
                   &\Gam^{abc}:\quad c=1, B=\gamma^a\gamma^{b,T}
                                            \gamma^c \qquad
                    \Gam^{ab9}:\quad c=1, B=0 \; .
       \end{alignat*}
Since after summation over the spin-polarizations, an antisymmetric matrix $B$
gives a vanishing contribution, the only non-vanishing terms for our
scattering process stem from $\Gam^{089}$ and $\Gam^{ijk};\, i,j,k=1,...,7$.
Hence we are able to reduce the second term to
       \begin{alignat*}{3}
                         &\sum_{s_1,s_2,s_3,s_4=1}^8
                                {\bar u}_{s_4}(p_4)
                                 \Gam^{ABC} 
                                 u_{s_3}(p_3)
                                 {\bar u}_{s_2}(p_2)
                                 \Gam_{ABC} 
                                 u_{s_1}(p_1)                       \\
                       =& \,3! \times E^2 \Bigg( -64 + 
                                \sum_{i<j<k}
                          \underbrace{\bigg[ \sum_{s_2,s_1}
                                         \left( \gamma^{ijk}
                                         \right)_{s_2s_1}
                                      \bigg]^2}_{(\pm8)^2}
                                \Bigg)                             \\
                       =& \,(3! \times 8E)^2  \; .
       \end{alignat*}
Putting the results for the two terms together, we arrive at the
following expression for the 3-form exchange amplitude
       \begin{equation}
              T_c = -\frac{\kappa^2}{\pi d}
                          \frac{128 (248)^2}{(3!)^2} 
                          s
                          \sum_{m \in \mathbb{Z}} (-1)^m
                          \frac{2s+3(p_m^{11})^2}
                               {-s+(p_m^{11})^2}   \; .
                  \label{cAmplitude}  
       \end{equation}
Using again (\ref{sum}) for the summation, we find
       \begin{equation*}
           \sum_{m \in \mathbb{Z}} (-1)^m
           \frac{2s+3(p_m^{11})^2}{-s+(p_m^{11})^2}   
          =3\sum_{m \in \mathbb{Z}} (-1)^m
           - 5\frac{\sqrt{s}d}
                    {\sin\left(\sqrt{s}d\right)}   \; .
       \end{equation*}
The first term describes an alternating sum, which does not converge
and requires some kind of regularization. In order to understand this
contribution, we will explore in a moment the $d\rightarrow \infty$ limit.
Therefore it proves useful
to express the above obtained amplitude in terms of the D=10
gravitational coupling constant $\kappa_{(10)}$ which is independent
of the compactification radius $R$ or $d$. Compactification
of M-theory on an $S^1$ of radius $R$ and a subsequent comparison of
its Einstein-Hilbert term with the Einstein-Hilbert term
coming from the effective action of the $D=10$ heterotic string
(in Einstein frame) leads to the following relationship between
the D=11 and the D=10 gravitational coupling constants
        \begin{equation}
            \kappa^2 = 2 d \kappa_{(10)}^2 \; .
            \label{KappaRelation}
        \end{equation}
Hence $T_c$ can be expressed as
        \begin{equation*}
            T_c = -\kappa_{(10)}^2
                        \frac{256 (248)^2}{(3!)^2 \pi}
                        s \left( 3 \sum_{m \in \mathbb{Z}} (-1)^m
                                -5 \frac{\sqrt{s}d}
                                        {\sin\left(\sqrt{s}d\right)}
                          \right)   \; .
        \end{equation*}
Since the first part of the amplitude, consisting of the alternating sum
and some $d$-independent prefactors, is independent of $d$, we can
equally well evaluate it at any $d$, in particular at $d\rightarrow
\infty$. Secondly, if we consider a large radius, the difference between
two adjacent values of $p_m^{11}$ becomes infinitesimally small and we are
allowed to replace the sum by an integral
        \begin{equation*}
           \lim_{d\rightarrow \infty}
           \sum_{m \in \mathbb{Z}}
           f( p_m^{11} = \frac{m}{R} = m\frac{\pi}{d} ) 
         = \lim_{d\rightarrow \infty}
           \frac{d}{\pi} 
           \int^\infty_{-\infty} d p^{11} f(p^{11}) \; .
        \end{equation*}  
Writing $(-1)^m = e^{ip_m^{11} d}$, we now encounter the following expression
for the alternating sum
        \begin{equation*}
           \sum_{m \in \mathbb{Z}} (-1)^m
         = \sum_{m \in \mathbb{Z}} e^{ip_m^{11} d}
         = \lim_{d\rightarrow \infty}
           \frac{d}{\pi} 
           \int^\infty_{-\infty} d p^{11} e^{ip^{11} d}
         = \lim_{d\rightarrow \infty}
           2 d \delta(d)
         = 0      \; .
        \end{equation*}
Thus finally the amplitude can be completely determined to be
        \begin{equation}
                        T_c (s)
                    =   \frac{\kappa^2}{\pi}
                        \frac{160 (248)^2}{9}
                        \frac{s^{3/2}}
                             {\sin\left(\sqrt{s}d\right)}
                        \; .
                          \label{FormFull}
        \end{equation}
The integration over the scattering angle from $0$ to $\pi/2$ is trivial
and results in
        \begin{equation}
                        {\cal T}_c (s)
                    =   \kappa^2
                        \frac{80 (248)^2}{9}
                        \frac{s^{3/2}}
                             {\sin\left(\sqrt{s}d\right)}
                        \; . \label{FormFullInteg}
        \end{equation}

\subsection{Two further Graviton exchange diagrams}
\begin{figure}[t]
              \begin{center}
               \begin{picture}(275,100)(-50,0)
                  \Gluon(-60,80)(-40,50){3}{4}
                  \Gluon(-60,20)(-40,50){3}{4}
                  \Photon(-40,50)(10,50){2}{8}
                  \Line(10,50)(30,80)
                  \Line(10,50)(30,20)
                  \Vertex(-40,50){1.0}
                  \Vertex(10,50){1.0}

                  \Text(-15,55)[b]{$h$}
                  \Text(-65,80)[r]{$B^c_{1C} \left(p_3\right)$}
                  \Text(-65,20)[r]{$B^d_{1D} \left(p_4\right)$}
                  \Text(35,80)[l]{$\lambda^a_2 \left(p_1\right)$}
                  \Text(35,20)[l]{$\lambda^b_2 \left(p_2\right)$}
               
                  \LongArrow(-40,18)(-40,38)
                  \LongArrow(10,18)(10,38)
                  \Text(-40,16)[t]{$x^{11}\!\!=\!0$}
                  \Text(10,16)[t]{$x^{11}\!\!=\!d$}

                  \Line(150,80)(170,50)
                  \Line(150,20)(170,50)
                  \Photon(170,50)(220,50){2}{8}
                  \Line(220,50)(240,80)
                  \Line(220,50)(240,20)
                  \Vertex(170,50){1.0}
                  \Vertex(220,50){1.0}

                  \Text(195,55)[b]{$h$}
                  \Text(145,80)[r]{$\lambda^c_1 \left(p_3\right)$}
                  \Text(145,20)[r]{$\lambda^d_1 \left(p_4\right)$}
                  \Text(245,80)[l]{$\lambda^a_2 \left(p_1\right)$}
                  \Text(245,20)[l]{$\lambda^b_2 \left(p_2\right)$}
               
                  \LongArrow(170,18)(170,38)
                  \LongArrow(220,18)(220,38)
                  \Text(170,16)[t]{$x^{11}\!\!=\!0$}
                  \Text(220,16)[t]{$x^{11}\!\!=\!d$}
               \end{picture}
               \caption{Vanishing Graviton exchange diagrams}

               \label{Picture5}

              \end{center}
            \end{figure}
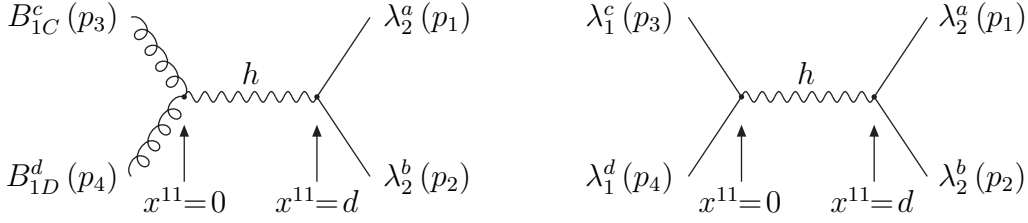

To complete our discussion of all relevant tree diagrams, which
contribute to a boundary-boundary interaction, we also have
to consider two further graviton exchange diagrams, depicted in
fig.\ref{Picture5}.
Both are, after performing the Wick-contractions, proportional to
         \begin{equation*}
            -{\bar u}_{s_2} (p_2)
             \left( \Gam^C \eta^{AB} - \Gam^B \eta^{AC}
             \right)
             u_{s_1} (p_1)
            +{\bar u}_{s_1} (p_1)
             \left( \Gam^C \eta^{AB} - \Gam^B \eta^{AC}
             \right)
             u_{s_2} (p_2)          \; ,
         \end{equation*}
which gives zero, if we do avail ourselves of (\ref{Bilinear1}).
Physically the vanishing of the diagrams is clear, since interchanging
the two gauginos of the final state gives a minus-sign, which
cannot be compensated for by the coupling to a graviton.
In the previously analysed case of the coupling between two
gauginos and the 3-form potential, the coupling delivers
an extra minus-sign under exchange of the two fermions, so that
the amplitude did not vanish in that case.

\subsection{Analysis of the amplitudes}
Gathering all the obtained amplitudes, integrated over the
scattering angle, we have
         \begin{alignat}{3}
               &{\cal T}_h (s) = -21 (248)^2 \kappa^2 
                      \frac{s^{3/2}}{\sin(\sqrt{s}d)}  
                         \label{FullAmp1} \\
               &{\cal T}_\psi (s) = i \frac{160}{3\pi} (248)^2 \kappa^2
                      \frac{s^{3/2}}{\sin(\sqrt{s}d)} 
                         \label{FullAmp2}  \\
               &{\cal T}_c (s) = \frac{80}{9} (248)^2 \kappa^2
                      \frac{s^{3/2}}{\sin(\sqrt{s}d)} 
                         \label{FullAmp3} \; .
         \end{alignat}
First of all, we have to determine the range of validity of
(\ref{FullAmp1})-(\ref{FullAmp3}). From the denominator we recognize
that singularities occur at the excitations of the Kaluza-Klein
states at $\sqrt{s}=m\pi/d=p_m^{11}\, ; \, m \in \mathbb{Z}$. Our
analysis did not cover contributions to the interaction amplitudes
coming from these states and only included the exchange of the massless
supergravity multiplet. Therefore, the range of validity of our
results is subjected to the following constraint, given by the first
Kaluza-Klein excitation
         \begin{equation*}
            0 \le \sqrt{s} < \frac{\pi}{d} \; .
         \end{equation*}
In the special case of vanishing CMS-energy $\sqrt{s}=0$, each amplitude
vanishes separately. This corresponds to the situation where the
boundary fields on each fixed plane run in parallel directions.
In this case we have trivially no interaction
between the two boundaries, as expected from the vanishing of the
susy-variations for this kinematics. In this special situation the flat
background with vanishing $G$-flux corresponds to a
stable ground state of the heterotic M-theory set-up. However, if there are
excitations on the boundary, by which we mean a kinematical situation
showing $\sqrt{s}>0$ for the boundary-fields, we see that
pure gravity leads to an attraction (since we are always below the first
Kaluza-Klein excitation energy),
whereas -- similar to the behaviour of the RR-forms in the analogous
D-brane case of Type II string theory -- the 3-form exchange leads to a
repulsion. If we choose the same CMS-energy for all three contributions,
then the attractive gravity dominates the weaker 3-form repulsion. Hence
the real part of the amplitudes indicates an instability which is caused
by an attractive force trying to bring the two boundaries closer together.

Thus the flat background with vanishing $G$-flux does not represent a stable
vacuum in the presence of arbitrary momenta of the boundary-fields.
An obvious guess as to the nature of a stable vacuum comes from the treatment
of heterotic M-theory compactified on a Calabi-Yau threefold \cite{Witten}.
There it has been shown, that with a non-vanishing $G$-flux on the Calabi-Yau
and the orbifold-direction, compactified heterotic M-theory exhibits a
warped-geometry. In view of the failure of the flat vacuum to represent a
stable configuration, one would naively think, that the warping of the
geometry should survive in the decompactification limit. Ten-dimensional
Poincar\'e-invariance indeed only allows for a non-trivial dependence on
$x^{11}$ and hence a warped-geometry. However, the very Poincar\'e-invariance
also requires $G_{KLMN}$ to vanish and therefore other sources for a warping
of space-time must be taken into account. 

The behaviour of the above calculated amplitudes is similar to the
weakly-coupled string-theory case in which an excited D-brane can decay into a
massless closed string state and the non-excited D-brane \cite{Klebanov}. Such 
a decay is also possible whenever the two massless waves
on the D-brane run in different directions and accordingly possess
$\sqrt{s}>0$. If the massless waves, however, run in the same direction,
i.e.~have $\sqrt{s}=0$, then one is dealing with a BPS state which does
not decay.

Curiously the gravitino exchange gives rise to an imaginary part. By
inspection of (\ref{GravitinoFull}) we find that the forward scattering
amplitude $T_\psi (s,\vartheta=0)$ is non-vanishing. Via the optical
theorem this would signal the opening of some inelastic channels for a
decay of an excited boundary and therefore an instability in a more
drastic sense. 

A last remark concerns unitarity. If we would evaluate total cross-sections
with the above amplitudes, then by integrating over the appropriate
phase space, we would get at high energies
         \begin{equation*}
            \sigma \sim | {\cal T}(s) |^2 s^2 
                   \sim \kappa^4 s^5 \; .
         \end{equation*}
However, unitarity of the S-matrix would lead for spinless states to the
following restriction on partial wave amplitudes
         \begin{equation*}
            \sigma_J \le \frac{P_J}{s^4} \; ,
         \end{equation*}
where $P_J$ is some polynomial in $J$ independent of $s$. Neglecting
$\vartheta$-dependent factors which arise for states with higher spin,
we conclude, that the total cross-section $\sigma$ which is the sum
of all $\sigma_J$, should decrease with increasing energy in order
to obey unitarity. Since our cross-sections increase with energy, they
violate unitarity. This is also plausible from the fact, that
Ho\v{r}ava-Witten supergravity is not gauge invariant at the classical
level and therefore no Ward-identities guarantee unitarity.
However, we have to keep in mind the restriction to the energy regime
$\sqrt{s} < \pi/d$ of our analysis. Should it happen, that
a violation of unitarity occurs at an energy much higher than $\pi/d$,
we would have to include the effects of the Kaluza-Klein excitations
to decide, whether unitarity is violated or obeyed.

\section{Comparison of the interaction amplitudes with the
         weakly coupled heterotic string amplitudes}
According to the conjecture made in \cite{HorWitt1}, we should recover
the D=10 weakly coupled heterotic $E_8\times E_8$ string theory in the
limit of small $R$ resp. $d$. Since the amplitudes which we have derived,
so far, describe the low-energy regime, we should also compare to the
analogous low-energy string amplitudes. Here we have to use the
expressions
(\ref{GravitonScattering1}),(\ref{GravitinoFull}),(\ref{FormFull})
which contain the full angular information.
In order to derive the zero radius limit,
we express all the derived amplitudes via (\ref{KappaRelation})
through the radius-independent $\kappa_{(10)}$ and then perform the limit
$\sqrt{s}d \rightarrow 0$
      \begin{alignat}{3}
           &T_h (s,\vartheta) = -\frac{4 \kappa_{(10)}^2}{\pi}
                        (248)^2
                        \frac{(25 s^2 - 32 tu)d}
                             {\sqrt{s}\sin\left(\sqrt{s}d\right)} 
           \quad
           \stackrel{\sqrt{s}d \rightarrow 0}{\longrightarrow}
           \quad
                       -\kappa_{(10)}^2
                        \frac{4 (248)^2}{\pi}
                        \left( 25s-32\frac{tu}{s} \right) 
                       \label{GravitonLim1}                    \\[3mm]
           &T_\psi (s,\vartheta) = i\frac{16 \kappa_{(10)}^2}
                                        {\pi}
                        (248)^2
                        \frac{\left( \left(s-t\right)\sqrt{-t}
                                    +\left(s-u\right)\sqrt{-u}
                              \right) d}
                             {\sin\left(\sqrt{s}d\right)}           \\
           &\phantom{T_\psi (s,\vartheta) =}
            \stackrel{\sqrt{s}d \rightarrow 0}{\longrightarrow}
                       i\frac{16 \kappa_{(10)}^2}
                                        {\pi}
                        (248)^2
                        \left( \left(s-t\right)\sqrt{-\frac{t}{s}}
                              +\left(s-u\right)\sqrt{-\frac{u}{s}}
                        \right)            
                       \label{GravitinoLim1}                   \\[3mm]
           &T_c (s) = \frac{320 \kappa_{(10)}^2}{9 \pi}
                        (248)^2
                        \frac{s^{3/2} d}
                             {\sin\left(\sqrt{s}d\right)} 
           \quad
           \stackrel{\sqrt{s}d \rightarrow 0}{\longrightarrow}
           \quad
                        \kappa_{(10)}^2
                        \frac{320 (248)^2}{9 \pi} s       \; .
                       \label{FormLim1}
      \end{alignat}
So far for the M-theory amplitudes.
Closed string amplitudes involve a factor $\kappa_{(10)}^{M-2+2L}$,
where $M$ is the number of external particles and $L$ the number
of loops. Hence with four external particles it is clear, that a factor
$\kappa_{(10)}^2$ corresponds to string tree-amplitudes as well.
Those heterotic string tree-amplitudes
can be found in \cite{GSW}\footnote{The
                                    translation between the momenta
                                    $k_i^{GSW}$ and
                                    Mandelstam-variables used by \cite{GSW}
                                    and the $p_i$ used in this paper is
                                    given by
                                    \begin{alignat*}{3}
                                        &k_1^{GSW} = p_1 \; , \qquad
                                         k_2^{GSW} = p_2 \; , \qquad
                                         k_3^{GSW} = -p_3 \; , \qquad
                                         k_4^{GSW} = -p_4 \; , \qquad    \\
                                        &s^{GSW} = -( k_1^{GSW}
                                                     +k_2^{GSW})^2
                                                 = -(p_1+p_2)^2 = s      \\
                                        &t^{GSW} = -( k_1^{GSW}
                                                     +k_4^{GSW})^2
                                                 = -(p_1-p_4)^2 = u      \\ 
                                        &u^{GSW} = -( k_1^{GSW}
                                                     +k_3^{GSW})^2
                                                 = -(p_1-p_3)^2 = t 
                                            \; .
                                    \end{alignat*}}. The terms,
which originate there from taking traces of four $E_8\times E_8$ group
generators $T_i$, must be discarded from our comparison, since they
correspond to processes where super Yang-Mills fields are exchanged
between the initial and final states.
What we want instead to compare to are the amplitudes
which are generated by the exchange of states of the supergravity
multiplet. Since they comprise singlet-representations under the
$E_8\times E_8$ gauge group, we merely encounter terms with
traces over two generators, respectively.
The string-theoretic tree-amplitudes adapted to our conventions
read
\begin{equation*}
       A = \kappa_{10}^2 K\left(\zeta_1,\frac{p_1}{2},
                                \zeta_2,\frac{p_2}{2},
                                \zeta_3,\frac{p_3}{2},
                                \zeta_4,\frac{p_4}{2}\right)
           C\left( s,t,u \right) G\left( p_1,p_2,p_3,p_4,
                                         T_1,T_2,T_3,T_4 \right) \; , 
\end{equation*}
where
\begin{alignat*}{3}
         &C\left( s,t,u \right) 
        = -\pi \frac{\Gam\left(-\frac{s}{8}\right)
                     \Gam\left(-\frac{t}{8}\right)
                     \Gam\left(-\frac{u}{8}\right)}
                    {\Gam\left(1+\frac{s}{8}\right)
                     \Gam\left(1+\frac{t}{8}\right)
                     \Gam\left(1+\frac{u}{8}\right)}            \\
         &G\left( p_1,p_2,p_3,p_4,T_1,T_2,T_3,T_4 \right)
        = \frac{1}{32}
          \left( 
          \frac{tu}{1+\frac{s}{8}}
          \text{tr}\left[ T_1T_2 \right]
          \text{tr}\left[ T_3T_4 \right]
          \right)              \; .
\end{alignat*}
The factor $G$
of \cite{GSW} also contains terms describing a t- and a u-channel exchange.
Since in the heterotic M-theory calculation for finite $d$, we get only
s-channel contributions for interactions of the boundary fields
via bulk fields, our expressions for $d\rightarrow 0$ should only
be compared to this very s-channel part of the string calculation.
For this reason we have omitted the t- and u-contributions to the
$G$-factor.
The generator $T_i$ corresponds to the $i^{\text{th}}$ external particle 
and tr is defined as the trace in the adjoint representation of
$E_8\times E_8$ divided by 30. The various $\zeta_i$ stand for the
polarization of the $i^{\text{th}}$ particle. If it is a spinor, we
have to substitute $\zeta_i = u_{s_i}(p_i)$, whereas for a gauge boson
we have to take its polarization $\zeta_i = \epsilon_i(p_i,\lambda_i)$.

The $K$-factor describes the kinematics of the interaction and is given
for the various cases\footnote{There is no factor $K(u_1,u_2,\epsilon_3,
                               \epsilon_4)$, since as in the heterotic
                               M-theory
                               calculation the $BB\rightarrow\lambda
                               \lambda$ contribution vanishes.}
by
\begin{alignat*}{3}
          &\phantom{-}K\left(\epsilon_1,\frac{p_1}{2},
                  \epsilon_2,\frac{p_2}{2},
                  \epsilon_3,\frac{p_3}{2},
                  \epsilon_4,\frac{p_4}{2}\right)             \\
        = &\phantom{-}\frac{1}{2^4}
           \bigg(
           -\frac{1}{4}\big( su\epsilon_1 \mydot \epsilon_3
                               \epsilon_2 \mydot \epsilon_4
                            +st\epsilon_2 \mydot \epsilon_3
                               \epsilon_1 \mydot \epsilon_4
                            +tu\epsilon_1 \mydot \epsilon_2
                               \epsilon_3 \mydot \epsilon_4
                       \big)                                  \\
          &-\frac{s}{2}\big( \epsilon_1 \mydot p_4 
                             \epsilon_3 \mydot p_2
                             \epsilon_2 \mydot \epsilon_4       
                            +\epsilon_2 \mydot p_3 
                             \epsilon_4 \mydot p_1
                             \epsilon_1 \mydot \epsilon_3
                            +\epsilon_1 \mydot p_3 
                             \epsilon_4 \mydot p_2
                             \epsilon_2 \mydot \epsilon_3
                            +\epsilon_2 \mydot p_4 
                             \epsilon_3 \mydot p_1
                             \epsilon_1 \mydot \epsilon_4
                       \big)                                  \\
          &+\frac{t}{2}\big(-\epsilon_1 \mydot p_2 
                             \epsilon_4 \mydot p_3
                             \epsilon_3 \mydot \epsilon_2       
                            -\epsilon_3 \mydot p_4 
                             \epsilon_2 \mydot p_1
                             \epsilon_1 \mydot \epsilon_4
                            +\epsilon_1 \mydot p_4 
                             \epsilon_2 \mydot p_3
                             \epsilon_3 \mydot \epsilon_4
                            +\epsilon_3 \mydot p_2 
                             \epsilon_4 \mydot p_1
                             \epsilon_1 \mydot \epsilon_2
                       \big)                                  \\  
          &+\frac{u}{2}\big(-\epsilon_2 \mydot p_1 
                             \epsilon_4 \mydot p_3
                             \epsilon_3 \mydot \epsilon_1       
                            -\epsilon_3 \mydot p_4 
                             \epsilon_1 \mydot p_2
                             \epsilon_2 \mydot \epsilon_4
                            +\epsilon_2 \mydot p_4 
                             \epsilon_1 \mydot p_3
                             \epsilon_3 \mydot \epsilon_4
                            +\epsilon_3 \mydot p_1 
                             \epsilon_4 \mydot p_2
                             \epsilon_2 \mydot \epsilon_1
                       \big)              
           \bigg)        
\end{alignat*}
\begin{equation*}
           K\left(u_1,\frac{p_1}{2},
                  \epsilon_2,\frac{p_2}{2},
                  u_3,\frac{p_3}{2},
                  \epsilon_4,\frac{p_4}{2}\right)
         = \frac{1}{2^3}
           \left(
           -\frac{u}{2}{\bar u}_1 \epsslash_2
            \left( \;\pslash_3 + \pslash_4 \right)
            \epsslash_4 u_3
           +\frac{s}{2}{\bar u}_1 \epsslash_4
            \left( \;\pslash_2 - \pslash_3 \right)
            \epsslash_2 u_3
            \right)      
\end{equation*}
\begin{alignat*}{3}
        &\phantom{-}K\left(\epsilon_1,\frac{p_1}{2},
                  u_2,\frac{p_2}{2},
                  u_3,\frac{p_3}{2},
                  \epsilon_4,\frac{p_4}{2}\right) 
        =           K\left(u_2,\frac{p_2}{2},
                  \epsilon_1,\frac{p_1}{2},
                  \epsilon_4,\frac{p_4}{2},
                  u_3,\frac{p_3}{2}\right)                    \\
        = &\phantom{-}
           \frac{1}{2^3}
           \left(
            \frac{u}{2}{\bar u}_2 \epsslash_1
            \left(\; \pslash_3 + \pslash_4 \right)
            \epsslash_4 u_3
           +s\left( {\bar u}_2 \epsslash_4 u_3 p_4 \mydot \epsilon_1
                   +{\bar u}_2 \epsslash_1 u_3 p_1 \mydot \epsilon_4
                   -{\bar u}_2 \pslash_4 u_3 \epsilon_1 \mydot \epsilon_4
             \right)
           \right)
           \qquad\qquad\qquad
\end{alignat*}
\begin{equation*}
           K\left(u_1,\frac{p_1}{2},
                  u_2,\frac{p_2}{2},
                  u_3,\frac{p_3}{2},
                  u_4,\frac{p_4}{2}\right)  
         = \frac{1}{2^2}
           \left(
           -\frac{s}{2} {\bar u}_2 \Gam^A u_3
                        {\bar u}_1 \Gam_A u_4
           +\frac{u}{2} {\bar u}_1 \Gam^A u_2
                        {\bar u}_4 \Gam_A u_3 
           \right)                               \; .
\end{equation*}
Summing over every occuring vectorial or spinorial polarization index,
we can simplify the kinematical factors further to
\begin{alignat*}{3}
          &\sum_{\lambda_1,\lambda_2,\lambda_3,\lambda_4}
      K\left(\epsilon_1,\frac{p_1}{2},
                  \epsilon_2,\frac{p_2}{2},
                  \epsilon_3,\frac{p_3}{2},
                  \epsilon_4,\frac{p_4}{2}\right)
        = -\frac{1}{8}\left(st+tu+us\right)
          +\frac{7}{16}\left(s^2+t^2+u^2\right)                  \\
          &\sum_{s_1,\lambda_2,s_3,\lambda_4}
      K\left(u_1,\frac{p_1}{2},
                  \epsilon_2,\frac{p_2}{2},
                  u_3,\frac{p_3}{2},
                  \epsilon_4,\frac{p_4}{2}\right)
         = 4s(s-u)\sqrt{-\frac{u}{s}}                            \\
          &\sum_{\lambda_1,s_2,s_3,\lambda_4}
      K\left(\epsilon_1,\frac{p_1}{2},
                  u_2,\frac{p_2}{2},
                  u_3,\frac{p_3}{2},
                  \epsilon_4,\frac{p_4}{2}\right)
         = 4s(s-t)\sqrt{-\frac{t}{s}}                            \\
          &\sum_{s_1,s_2,s_3,s_4}
      K\left(u_1,\frac{p_1}{2},
                  u_2,\frac{p_2}{2},
                  u_3,\frac{p_3}{2},
                  u_4,\frac{p_4}{2}\right)  
         = -4 \left( 3 s^2 + (t-u)^2 \right) \; .
\end{alignat*}
In the low-energy limit $\alpha' s,\alpha' t,\alpha' u \rightarrow 0$ 
we have
\begin{equation*}
     C\left( s,t,u \right) \rightarrow \frac{2^9 \pi}{stu}
\end{equation*}
such that finally we arrive at the following expressions for the low-energy
limit of the heterotic string amplitudes
\begin{alignat}{3}
         A_{BB\stackrel{s}{\rightarrow}BB} 
      &= \pi\kappa_{10}^2 16 
         \left(\text{tr}\left[ T_1T_2 \right]\right)^2
         \left( s - \frac{tu}{s} \right)  
              \label{GravitonLim2}                        \\
         A_{\lambda B\stackrel{s}{\rightarrow}\lambda B}
        +A_{\lambda B\stackrel{s}{\rightarrow}B \lambda}
      &= \pi\kappa_{10}^2 4\times 16
         \left(\text{tr}\left[ T_1T_2 \right]\right)^2
         \left( (s-t)\sqrt{-\frac{t}{s}}
               +(s-u)\sqrt{-\frac{u}{s}}
         \right)                          
              \label{GravitinoLim2}                       \\
         A_{\lambda\lambda\stackrel{s}{\rightarrow}\lambda\lambda} 
      &= -\pi\kappa_{10}^2 (16)^2
         \left(\text{tr}\left[ T_1T_2 \right]\right)^2
         \left( s - \frac{tu}{s} \right)             \; .
              \label{FormLim2}
\end{alignat}
If we compare these with
(\ref{GravitonLim1}),(\ref{GravitinoLim1}),(\ref{FormLim1}), we recognize
substantial differences. Whereas (\ref{GravitonLim1}) and
(\ref{GravitonLim2}) deviate mildly in their functional dependence
on $s,t,u$, the discrepancy between (\ref{FormLim1}) and
(\ref{FormLim2}) is manifest. The string amplitude shows an angular
dependence but the heterotic M-theory amplitude is isotropic.
The gravitino exchange amplitudes agree in their angular dependence.
Nevertheless the string amplitude is real, whereas its M-theoretic
counterpart is purely imaginary.

However, such a disagreement had to be expected.
Generally, a low-energy description in terms of effective
supergravity is only valid at large distances resp. small curvatures.
Furthermore, Ho\v{r}ava-Witten supergravity is organized as a long
wavelength expansion in the parameter $\kappa^{2/3}$, assumed to be small as
compared to the eleventh-dimensional Planck-scale.
However, in the limit $d \rightarrow 0$ of coinciding boundaries, the
long-wavelength supergravity approximation breaks down and one cannot
trust the order $\kappa^{2/3}$ expansion any longer. Therefore the
effective description gives incorrect results.

\section{Summary and Conclusion}
Our aim in this paper has been to vet whether heterotic M-theory on a flat
background in its concrete low-energy formulation as Ho\v{r}ava-Witten
supergravity is stable.
Therefore, we used the background field method to derive interaction
amplitudes between the fields living on the boundaries. Each amplitude
vanishes trivially for non-excited boundaries which corresponds to a
BPS ground state configuration as in the similar D-brane case.
Nevertheless for a gauge excitation on the boundary with a CMS-energy
$0< \sqrt{s} < \pi/d$ below the first Kaluza-Klein resonance, the
gravitational attraction
dominated the 3-form repulsion, thereby giving rise to a net attractive
force between the two boundaries. If one included also the contribution
from the gravitino exchange, this generated an imaginary part for
the amplitude. Via the optical theorem, an even more drastic instability
of the effective theory was signalled. The consequence would be an
inelastic decay of the flat vacuum set-up towards an energetically more
favoured configuration. To avoid this instability, we argued, that already for
the uncompactified heterotic M-theory the vacuum should exhibit a
warped-geometry. 

We briefly indicated that unitarity seems to be violated by the interaction
amplitudes, which is
also understandable from the fact, that the effective Ho\v{r}ava-Witten
description is classically not gauge invariant.
Since our treatment was reliable only for energies below the first
Kaluza-Klein resonance, a seeming violation of unitarity above that
limit could not be tested.

Finally, we extrapolated the M-theory amplitudes to the extreme limit in which
both boundaries coincide. According to the conjecture, this is the
realm where we should recover the $E_8\times E_8$ heterotic string.
But a comparison with the low-energy heterotic string amplitudes
showed no complete agreement. However, this discrepancy had to be expected,
since in this limit the
effective supergravity description and in particular the long wavelength
expansion of Ho\v{r}ava-Witten supergravity breaks down and leads to false
results.

\bigskip
\noindent {\large \bf Acknowledgements}\\[2ex] 
We would like to thank I.~Antoniadis, M.~Faux, P.~Ho\v{r}ava, and
D.~L\"{u}st for helpful comments or remarks on partial aspects of this work.

\newpage
\begin{appendix}

\section{Appendix}
\subsection{Notations}
       \noindent
       \underline{Tensors}:
       \begin{alignat*}{3}
             A_{(M_1 ... M_n)} &= \frac{1}{n!}\left(
             A_{M_1 ... M_n}\pm (n!-1) \text{ symmetric permutations} 
                                       \right)                     \\
             A_{[M_1 ... M_n]} &= \frac{1}{n!}\left(
             A_{M_1 ... M_n}\pm (n!-1) \text{ antisymmetric permutations} 
                                       \right)                   
       \end{alignat*}

\subsection{Mandelstam-Variables and Kinematic for the CMS}
   \label{AppMandelstam} 
   \underline{Center-of-Mass variables}:  
        \begin{equation}
           \text{scattering angle: } 0 \le \vartheta \le \pi \qquad
           \text{CMS-Energy: } E 
        \end{equation}  
    Without loss of generality we can arrange the scattering such that
    the two incoming fields with momenta $p_1,p_2$ collide head-on
    in the CMS-system in the direction of the
    $9^{th}$ coordinate axis and the outgoing fields move on the plane
    spanned by the $8^{th}$ and $9^{th}$ coordinate axes.                 
    We choose $\vartheta$ to be the angle between
    $p_1$ and $p_3$.
    Concretely we take for the D=10 momenta
    \begin{alignat*}{3}
        &p_1 = \left(\frac{E}{2},0,...,0,\frac{E}{2}\right)
               \; , \quad
         p_2 = \left(\frac{E}{2},0,...,0,-\frac{E}{2}\right)
               \; ,                                    \\
        &p_3 = \left(\frac{E}{2},0,...,0,\frac{E}{2}\sin\vartheta,
                                         \frac{E}{2}\cos\vartheta\right)
               \; , \quad
         p_4 = \left(\frac{E}{2},0,...,0,-\frac{E}{2}\sin\vartheta,
                                         -\frac{E}{2}\cos\vartheta\right)
                    \; .
    \end{alignat*}  
    \noindent
    \underline{Mandelstam-Variables}:
        \begin{alignat*}{3}
           &s = -\left(p_1+p_2\right)^2 
              = -\left(p_3+p_4\right)^2 
              = E^2                                  \\
           &t = -\left(p_1-p_3\right)^2 
              = -\left(p_2-p_4\right)^2
              = -\frac{s}{2}\left[ 1 - \cos\vartheta \right]
              = -s \sin^2 \frac{\vartheta}{2}
              \le 0                                  \\
           &u = -\left(p_1-p_4\right)^2 
              = -\left(p_2-p_3\right)^2
              = -\frac{s}{2}\left[ 1 + \cos\vartheta \right]
              = -s \cos^2 \frac{\vartheta}{2}
              \le 0                                  \\
           &\qquad\qquad\qquad\qquad\qquad s+t+u = 0
        \end{alignat*}

\subsection{D=10 polarization vectors}
    \noindent
    \underline{CMS-Polarization Vectors}: Let the
     D=10 momentum in the CMS be
          \begin{equation}
               p= \frac{E}{2} \left( 1, ... , \sin\vartheta, 
                                    \cos\vartheta 
                              \right) \; . 
          \end{equation}
    Then we have 8 transverse polarizations, which are given in the
    CMS by the following real vectors
          \begin{alignat}{3}
               &\epsilon(p,1) = (0,1,0,...,0)                         \\
               &\epsilon(p,2) = (0,0,1,...,0)                         \\
               &\qquad \vdots                                         \\
               &\epsilon(p,7) = (0,0,0,...,1,0,0)                     \\
               &\epsilon(p,8) = (0,0,0,...,0,\cos\vartheta,
                                            -\sin\vartheta)
          \end{alignat}

    \noindent
    \underline{Useful Contractions}:
          If we sum over all polarizations, we get
          \begin{equation}
               \sum_{\lambda =  1}^8 
                p_{1,A} \epsilon^A \!\left( p_2,\lambda \right)
              = \frac{E}{2} \sin\left( \vartheta_1 - \vartheta_2 \right)
               \; .
          \end{equation}    

          \noindent
          Contractions of two polarization vectors are given by
          \begin{equation}
              \sum_{\lambda,\tilde{\lambda}=1}^8 
              \epsilon_A \!\left(p,\lambda\right)
              \epsilon^A \!\left(\tilde{p},\tilde{\lambda}\right)
            = 7 + \cos\left( \vartheta - \tilde{\vartheta} \right)
                                 \; .
          \end{equation}

\subsection{D=11 Gamma-matrices and D=10 Dirac-spinors}
         \label{AppGamma}
         We take the D=11 $SO(1,10)$ spin $32\times 32$ matrices to be in a
         real Majorana-representation
          \begin{alignat*}{3}
             \Gam^0 &= -i\sigma_2 \otimes I_{16}
                     = \left( \begin{array}{cc} 0 & -1 \\
                                                1 & 0
                              \end{array} 
                       \right)        
                       \otimes
                       \left( \begin{array}{cc} I_8 & 0 \\
                                                0 & I_8
                              \end{array} 
                       \right)                                        \\
             \Gam^a &= \sigma_1 \otimes \gamma^a_{16}
                     = \left( \begin{array}{cc} 0 & 1 \\
                                                1 & 0
                              \end{array} 
                       \right)        
                       \otimes
                       \left( \begin{array}{cc} 0 & \gamma^a_8 \\
                                                \gamma^{a,T}_8 & 0
                              \end{array} 
                       \right)
                       \; ;  \quad a=1,...,8                          \\
             \Gam^9 &= -\sigma_1 \otimes \gamma^9_{16} 
                     = \left( \begin{array}{cc} 0 & -1 \\
                                                -1 & 0
                              \end{array} 
                       \right)        
                       \otimes
                       \left( \begin{array}{cc} I_8 & 0 \\
                                                0 & -I_8
                              \end{array} 
                       \right)                                        \\
                                                                      \\
          \Gam^{10} &\equiv \Gam^0\Gam^1...\Gam^8\Gam^9               \\[3mm]
                    &= \sigma_3 \otimes I_{16} 
                     = \left( \begin{array}{cc} 1 & 0 \\
                                                0 & -1
                              \end{array} 
                       \right)        
                       \otimes
                       \left( \begin{array}{cc} I_8 & 0 \\
                                                0 & I_8
                              \end{array} 
                       \right) 
                     = \left( \begin{array}{cc} I_{16} & 0 \\
                                                0 & -I_{16}
                              \end{array} 
                       \right)
          \end{alignat*}
where we use $\epsilon_{0...9} = 1 = -\epsilon^{0...9}$. The 
Dirac-matrices satisfy
          \begin{equation*}
               \big\{ \Gam^M, \Gam^N \big\}
             = 2\eta^{MN} = 2 (-,+,...,+) \; ,
          \end{equation*}
while the real
$16\times 16$ $\gamma^a_{16}$ submatrices obey the relations        
          \begin{alignat*}{3}
              &\big\{ \gamma^a_{16}, \gamma^b_{16} \big\}
             = 2\delta^{ab}
               \;, \qquad
               \gamma^{a,T}_{16}
             = \gamma^a_{16}
               \;, \qquad
                \big( \gamma^a_{16} \big)^2 = I_{16}                 \\
              &\gamma^9_{16} \equiv
               \gamma^1_{16}...\gamma^8_{16}
             = \left( \begin{array}{cc} I_8 & 0 \\
                                        0 & -I_8
                      \end{array} 
               \right)
               \;, \qquad           
               \gamma^{9,T}_{16}
             = \gamma^9_{16}
               \;, \qquad
               \big( \gamma^9_{16} \big)^2 = I_{16}  \; .               
          \end{alignat*}
Finally the $8\times 8$ submatrices $\gamma^a$ are defined as
          \begin{alignat*}{3}
              &\gamma^1 = i\sigma_2 \otimes i\sigma_2 \otimes
                          i\sigma_2
                          \; , \quad
               \gamma^2 = I_2 \otimes \sigma_1 \otimes
                          i\sigma_2                                 \\
              &\gamma^3 = I_2 \otimes \sigma_3 \otimes
                          i\sigma_2
                          \; , \quad
               \gamma^4 = \sigma_1 \otimes i\sigma_2 \otimes
                          I_2                                       \\
              &\gamma^5 = \sigma_3 \otimes i\sigma_2 \otimes
                          I_2
                          \; , \quad
               \gamma^6 = i\sigma_2 \otimes I_2 \otimes
                          \sigma_1                                  \\
              &\gamma^7 = i\sigma_2 \otimes I_2 \otimes
                          \sigma_3
                          \; , \quad
               \gamma^8 = I_2 \otimes I_2 \otimes
                          I_2    
                        = I_8
          \end{alignat*}
and satisfy
          \begin{equation*}
              \gamma^a \gamma^{b,T}
             +\gamma^b \gamma^{a,T} = 2 \delta^{ab}
              \; , \quad
              \gamma^{i,T}=-\gamma^i\; ; i=1,...,7
              \; , \quad
              \gamma^{8,T}=-\gamma^8 \; .
          \end{equation*}

\noindent
\underline{D=10 Weyl-spinor}:
We have to deal with ten-dimensional Majorana-Weyl spinors for the gauginos
with positive chirality only. For the special ten-dimensional momentum
$p_1 = \frac{E}{2}(1,0,...,0,1)$, we find from the Dirac-equation
$\Gamma^A \partial_A \lambda(x) = 0$, the following spinor expression
in momentum space
          \begin{equation*}
               u_s(p_1) = \sqrt{N}
                          \left( \begin{array}{c} 
                                                  0 \\
                                                  e_s \\
                                                  0  \\
                                                  0
                                 \end{array} 
                          \right)  \; ,
          \end{equation*} 
where $e_s; s=1,...,8$ denotes the $\text{s}^{th}$ unit vector.
In our calculation we actually need the slightly more general
spinor corresponding to the ten-dimensional momentum
$p=\frac{E}{2}(1,0,...,0,\sin\vartheta,\cos\vartheta)$. We can generate
$p$ from $p_1$
by a rotation in the 8-9 plane
          \begin{equation*}
            p^A = R_\vartheta p_1^A
                = \left( \begin{array}{ccc}
                               \ddots & 0 & 0 \\
                               0 & \cos\vartheta & \sin\vartheta \\
                               0 & -\sin\vartheta & \cos\vartheta
                         \end{array} 
                  \right)   
                  \left( \begin{array}{c} 
                                          p_1^0 \\
                                          \vdots \\
                                          p_1^8 \\
                                          p_1^9
                         \end{array} 
                  \right) \; .
          \end{equation*} 
The corresponding action on the spinor $u_s(p_1)$ is given by
          \begin{alignat}{3}
             u_s(p) &= e^{\frac{\vartheta}{2}\Gam^8\Gam^9} u_s(p_1)
                     = \left( \cos\left(\frac{\vartheta}{2}\right) 
                              I_{32}
                             +\sin\left(\frac{\vartheta}{2}\right) 
                              \Gam^8 \Gam^9
                       \right) u_s(p_1)                   \notag     \\ 
                    &= \sqrt{N} \left( \begin{array}{c} 
                            \sin\left(\frac{\vartheta}{2}\right) e_s \\
                            \cos\left(\frac{\vartheta}{2}\right) e_s \\
                                                                   0 \\
                                                                   0
                                       \end{array} 
                                \right) \; .
                        \label{GauginoSpinor}
          \end{alignat}
As a convenient normalization choice we choose
         \begin{equation*}
             N \equiv E \; .
         \end{equation*}
The charge conjugation matrix $C_{\alpha\beta}$ will be taken as
         \begin{equation*}
             C_{\alpha\beta} 
           = \left(\Gam^0\right)^\alpha_{\phantom{\alpha}\beta} \; , \quad
             C^{\alpha\beta}
           = \left( \Gam^{0,-1} \right)^\alpha_{\phantom{\alpha}\beta}
           = \left( \Gam^{0,T} \right)^\alpha_{\phantom{\alpha}\beta}
           = -\left( \Gam^0 \right)^\alpha_{\phantom{\alpha}\beta} \; .
         \end{equation*}

\noindent
\underline{Symmetry properties of Bilinears}:\\
For arbitrary momenta $p$ and $p'$ one obtains
          \begin{alignat}{3}
               &{\bar u}_s(p) \Gam^A u_{s'}(p')
              = {\bar u}_{s'}(p') \Gam^A u_{s}(p)
                    \label{Bilinear1}                                 \\
               &{\bar u}_s(p) \Gam^{ABC} u_{s'}(p')
              =-{\bar u}_{s'}(p') \Gam^{ABC} u_{s}(p)                 \\
               &{\bar u}_s(p) \Gam^{ABCDE} u_{s'}(p')
              ={\bar u}_{s'}(p') \Gam^{ABCDE} u_{s}(p) \; .  
          \end{alignat}

\subsection{Hyperplane Gauge Field Operators}
        \label{Operators}
        \underline{Fourier-Decomposition of the Field Operators}:
        \begin{alignat}{3}
            &B^a_A (x) 
           = \int \frac{d^9 k}{\left(2\pi\right)^9 2 k^0}
             \sum_{\lambda=1,...,8}
             \epsilon_A \!\left(k,\lambda\right)
             \left[ b^a_\lambda \left(k\right)
                    e^{ikx}
                   +b^{a,\dagger}_\lambda \left( k \right)
                    e^{-ikx}
             \right]                                                  \\
            &\lambda^a (x) 
           = \int \frac{d^9 k}{\left(2\pi\right)^9 2 k^0}
             \sum_{s=1,...,8}
             u_s(k)
             \left[ d^a_s(k) e^{ikx}
                   +d^{a,\dagger}_s (k) e^{-ikx}
             \right]                                                  \\
        \end{alignat} 

        \noindent
        \underline{Anti-/Commutators}:
        \begin{alignat}{3}
            &\big[ b^a_\lambda (k),
                   b^{b,\dagger}_{\lambda'} (k')
             \big]
           = \delta^9 (k-k') 
             \delta_{\lambda\lambda'}
             \delta^{ab} (2\pi)^9 2 k^0                               \\
            &\big[ b^a_\lambda (k),
                   b^b_{\lambda'} (k')
             \big]
           = \big[ b^{a,\dagger}_\lambda (k),
                   b^{b,\dagger}_{\lambda'} (k')
             \big]
           = 0                                                        \\
           &\big\{ d^a_s(k), d^{b,\dagger}_{s'}(k') \big\}
           = \delta^9 (k-k') 
             \delta_{ss'}
             \delta^{ab} (2\pi)^9 2 k^0                               \\
            &\big\{ d^a_s(k), d^b_{s'}(k')
             \big\}
           = \big\{ d^{a,\dagger}_s(k), b^{b,\dagger}_{s'}(k')
             \big\}
           = 0 
        \end{alignat}

        \noindent
        \underline{Wick-Contractions}:
        \begin{alignat}{3}
            \thinlines
            \put(2.7,-5){\line(1,0){29.3}}
            \put(2.7,-5){\line(0,1){3}}
            \put(32,-5){\line(0,1){3}}
            &b^a_\lambda \!\left(p\right) B^b_A \left(x\right)
          = \delta^{ab} \epsilon_A \!\left(p,\lambda\right) e^{-ipx} 
                           \label{Wick1}                              \\
            \thinlines
            \put(4,-5){\line(1,0){32.5}}
            \put(4,-5){\line(0,1){3}}
            \put(36.5,-5){\line(0,1){3}}
            &B^a_A \!\left(x\right) b^{b,\dagger}_\lambda
                   \left(p\right) 
          = \delta^{ab} \epsilon_A \!\left(p,\lambda\right) e^{ipx}   
                           \label{Wick2}                              \\
            \thinlines
            \put(2,-5){\line(1,0){29}}
            \put(2,-5){\line(0,1){3}}
            \put(31,-5){\line(0,1){3}}
           &d^a_s \!\left(p\right) \lambda^{b\alpha} (x) 
          = \delta^{ab} u^\alpha_s \left(p\right) e^{-ipx}   
                           \label{Wick3}                              \\
            \thinlines
            \put(3,-5){\line(1,0){30.5}}
            \put(3,-5){\line(0,1){3}}
            \put(33.5,-5){\line(0,1){3}}
           &\lambda^{a\alpha} \!(x) d^{b,\dagger}_s (p)      
          = \delta^{ab} u^\alpha_s \left(p\right) e^{ipx}
                           \label{Wick4}     
        \end{alignat}
\end{appendix}

 \newcommand{\zpc}[3]{{\sl Z. Phys.} {\bf C#1} (19#2) #3}
 \newcommand{\zp}[3]{{\sl Z. Phys.} {\bf #1} (19#2) #3}
 \newcommand{\npb}[3]{{\sl Nucl. Phys.} {\bf B#1} (19#2)~#3}
 \newcommand{\plb}[3]{{\sl Phys. Lett.} {\bf B#1} (19#2) #3}
 \newcommand{\prd}[3]{{\sl Phys. Rev.} {\bf D#1} (19#2) #3}
 \newcommand{\prl}[3]{{\sl Phys. Rev. Lett.} {\bf #1} (19#2) #3}
 \newcommand{\prep}[3]{{\sl Phys. Rep.} {\bf #1} (19#2) #3}
 \newcommand{\fp}[3]{{\sl Fortschr. Phys.} {\bf #1} (19#2) #3}
 \newcommand{\nc}[3]{{\sl Nuovo Cimento} {\bf #1} (19#2) #3}
 \newcommand{\ijmp}[3]{{\sl Int. J. Mod. Phys.} {\bf #1} (19#2) #3}
 \newcommand{\rmp}[3]{{\sl Rev. Mod. Phys.} {\bf #1} (19#2) #3}
 \newcommand{\ptp}[3]{{\sl Prog. Theor. Phys.} {\bf #1} (19#2) #3}
 \newcommand{\sjnp}[3]{{\sl Sov. J. Nucl. Phys.} {\bf #1} (19#2) #3}
 \newcommand{\cpc}[3]{{\sl Comp. Phys. Commun.} {\bf #1} (19#2) #3}
 \newcommand{\mpla}[3]{{\sl Mod. Phys. Lett.} {\bf A#1} (19#2) #3}
 \newcommand{\cmp}[3]{{\sl Commun. Math. Phys.} {\bf #1} (19#2) #3}
 \newcommand{\jmp}[3]{{\sl J. Math. Phys.} {\bf #1} (19#2) #3}
 \newcommand{\nim}[3]{{\sl Nucl. Instr. Meth.} {\bf #1} (19#2) #3}
 \newcommand{\el}[3]{{\sl Europhysics Letters} {\bf #1} (19#2) #3}
 \newcommand{\ap}[3]{{\sl Ann. of Phys.} {\bf #1} (19#2) #3}
 \newcommand{\jetp}[3]{{\sl JETP} {\bf #1} (19#2) #3}
 \newcommand{\jetpl}[3]{{\sl JETP Lett.} {\bf #1} (19#2) #3}
 \newcommand{\acpp}[3]{{\sl Acta Physica Polonica} {\bf #1} (19#2) #3}
 \newcommand{\vj}[4]{{\sl #1~}{\bf #2} (19#3) #4}
 \newcommand{\ej}[3]{{\bf #1} (19#2) #3}
 \newcommand{\vjs}[2]{{\sl #1~}{\bf #2}}
 \newcommand{\hep}[1]{{\sl hep--ph/}{#1}}
 \newcommand{\desy}[1]{{\sl DESY-Report~}{#1}}

\end{document}